\documentclass[a4paper,10pt]{revtex4}
\usepackage{graphicx} 
\usepackage{amsmath} 
\usepackage{amssymb} 
\usepackage{bm} 
\usepackage{dcolumn}
\usepackage{color}
\usepackage{mathrsfs}
\usepackage{amsfonts}
\usepackage{varioref}
\usepackage{mathrsfs}
\usepackage{graphicx}
\usepackage{latexsym}
\usepackage{amsmath}
\usepackage{amssymb}
\usepackage{textcomp}
\usepackage{amsbsy}
\usepackage{graphics}
\usepackage{epstopdf}
\usepackage{color}
\usepackage[caption=false]{subfig}

\RequirePackage[colorlinks,citecolor=blue,urlcolor=magenta,linkcolor=blue]{hyperref}
\input epsf

\allowdisplaybreaks[4]
\newcommand{\e}{\mathrm{e}}

\begin{document}

\tolerance=5000

\title{Holographic realization of constant roll inflation and dark energy: An unified scenario}

\author{Shin'ichi~Nojiri$^{1,2}$\,\thanks{nojiri@gravity.phys.nagoya-u.ac.jp},
Sergei~D.~Odintsov$^{3,4}$\,\thanks{odintsov@ieec.uab.es},
Tanmoy~Paul$^{5}$\,\thanks{pul.tnmy9@gmail.com}} \affiliation{
$^{1)}$ Department of Physics, Nagoya University,
Nagoya 464-8602, Japan \\
$^{2)}$ Kobayashi-Maskawa Institute for the Origin of Particles
and the Universe, Nagoya University, Nagoya 464-8602, Japan \\
$^{3)}$ ICREA, Passeig Luis Companys, 23, 08010 Barcelona, Spain\\
$^{4)}$ Institute of Space Sciences (ICE, CSIC) C. Can Magrans s/n, 08193 Barcelona, Spain\\
$^{5)}$ National Institute of Technology Jamshedpur, Department of Physics, Jamshedpur - 831 014, India.}


\tolerance=5000

\begin{abstract}
In the formalism of generalized holographic dark energy, the infrared cut-off $L_\mathrm{IR}$ is generalized to the form,
$L_\mathrm{IR} = L_\mathrm{IR}\!\left(L_\mathrm{p}, \dot L_\mathrm{p}, \ddot L_\mathrm{p}, \cdots, L_\mathrm{f}, \dot L_\mathrm{f},
\ddot L_\mathrm{f}, \cdots, a, H, \dot H, \ddot H, \cdots\! \right)$, where $L_\mathrm{p}$ and $L_\mathrm{f}$
are the particle horizon and the future horizon, respectively (moreover, $a$ is the scale factor and $H$ is the Hubble parameter of the universe).
Based on such formalism, we establish a holographic realization of constant roll inflation during the early universe,
where the corresponding cut-off depends on the Hubble parameter and its derivatives (up to the second order).
The viability of this holographic constant roll inflation with respect to the Planck data in turn puts a certain bound
on the infrared cut-off at the time of horizon crossing.
Such holographic correspondence of constant roll inflation is extended to the scenario where the infrared cut-off is corrected
by the ultraviolet one, which may originate due to quantum effects.
Besides the mere inflation, we further propose the holographic realization of an $unified$ cosmic scenario from constant roll inflation (at the early time)
to the dark energy era (at the late time) with an intermediate radiation dominated era followed by a Kamionkowski like reheating stage.
In such a unified holographic scenario, the inflationary quantities (like the scalar spectral index
and the tensor-to-scalar ratio) and the dark energy quantities (like the dark energy EoS parameter and the present Hubble rate)
prove to be simultaneously compatible with observable constraints for suitable ranges of the infrared cut-off and the other model parameters. Moreover the curvature perturbations at super-Hubble scale prove to be a constant (with time) during the entire cosmic era, which in turn ensures the stability of the model under consideration.
\end{abstract}

\maketitle

\section{Introduction}\label{SecI}

One of the most important puzzles in today's cosmology is why the universe undergoes an {\it accelerating} phase
at the two extreme curvature regimes, i.e., during the early time and during the late time.
Several higher curvature gravity theories (like $f(R)$ theory, the Gauss-Bonnet gravity theory, etc.) have been used to resolve this issue,
and well, they earned quite a success in this direction (see \cite{Capozziello:2011et, Nojiri:2010wj, Nojiri:2017ncd} for extensive reviews).
In the higher curvature theories, the Einstein-Hilbert action is suitably generalized, which may originate from the diffeomorphism property
of the gravitational action or from the string theory.
One other arena of string theory is the holographic principle that particularly originates from black hole thermodynamics and string theory
and establishes a connection of the infrared cutoff of a quantum field theory, which is related to the vacuum energy, with the largest distance
of this theory \cite{tHooft:1993dmi, Susskind:1994vu, Witten:1998qj, Bousso:2002ju}.
Such holographic consideration is currently running through its full pace in the field of cosmology.
In particular, holographic dark energy (HDE), which is based on the holographic principle rather than adding some extra term
in the Lagrangian of matter, proves to successfully describe the late time acceleration era
of the universe \cite{Li:2004rb, Li:2011sd, Wang:2016och, Pavon:2005yx, Nojiri:2005pu, Zhang:2005yz, Elizalde:2005ju,Gong:2004cb,Malekjani:2012bw,Khurshudyan:2016gmb,
Gao:2007ep,Li:2008zq,Anagnostopoulos:2020ctz,Li:2009bn,Feng:2007wn,Sheykhi:2023woy,Sheykhi:2022jqq,Lu:2009iv,Huang:2004wt,Mukherjee:2017oom,Nojiri:2017opc,
Saridakis:2020zol, Barrow:2020kug, Adhikary:2021xym, Srivastava:2020cyk, Nojiri:2022nmu, Bhardwaj:2021chg, Chakraborty:2020jsq}. Note that
generalized holographic dark energy introduced in \cite{Nojiri:2005pu} gives all known HDEs as particular cases -- this was explicitly proven in \cite{Nojiri:2021iko, Nojiri:2021jxf}. Besides dark energy, the holographic principle has also been used to realize inflation during the early phase of the universe,
particularly the slow-roll inflation \cite{Nojiri:2019kkp} (see also the subsequent papers \cite{Paul:2019hys, Bargach:2019pst, Elizalde:2019jmh, Oliveros:2019rnq, Chakraborty:2020tge}). Moreover, the holographic inflation with a suitable cut-off seems to be compatible with the recent Planck data.
Actually, during the early universe the holographic energy density, which is inversely proportional to the squared infrared cut-off of the theory,
becomes large and hence can drive inflation.
Recently we proposed a unified cosmological scenario of slow-roll inflation and the dark energy era from a holographic point of view
in a covariant way \cite{Nojiri:2020wmh}.
The holographic realization to describe the early universe is even extended to the bouncing cosmology, in which case,
the holographic energy density violates the null energy condition and triggers a non-singular bouncing universe \cite{Nojiri:2019yzg, Brevik:2019mah}.

Regarding holographic inflation, as we have mentioned earlier that the holographic principle has been used particularly for {\it slow-roll}
inflation \cite{Nojiri:2019kkp, Paul:2019hys, Bargach:2019pst, Elizalde:2019jmh, Oliveros:2019rnq, Chakraborty:2020tge, Nojiri:2020wmh}.
In the slow-roll regime, an almost flat scalar potential is generally considered that leads to a negligible acceleration of the scalar field under consideration.
In particular, the term $\ddot{\phi}$ (where $\phi$ is the scalar field and an overdot denotes the derivative with respect to cosmic time)
is negligible with respect to the Hubble friction and the restoring force.
As a qualitatively different condition, people considered the constant roll inflation where the scalar field rolls at a constant rate, in particular,
$\ddot{\phi}/\left(H\dot{\phi}\right) = \beta$ (with $\beta$ is an arbitrary constant and not necessarily that $\beta \ll 1$) \cite{Motohashi:2014ppa}.
Clearly, in the regime of $\beta \ll 1$, the constant roll inflation reduces to the standard slow-roll one.
A lot of successful attempts have been made to propose constant roll inflation during the early universe in the realm of various modified gravity models,
which are also consistent with the observable data (for instance, see \cite{Kinney:2005vj, Kumar:2015mfa, Cook:2015hma, Nojiri:2017qvx, Gao:2017uja,
Odintsov:2017qpp, Odintsov:2017hbk, Cicciarella:2017nls, Ito:2017bnn, Mohammadi:2018oku, Elizalde:2018now}, but not limited to).
Due to the considerable successes of constant roll inflation and the holographic principle, it becomes important to examine whether any connection
exists between them, and we would like to mention that this still demands a proper investigation.

Coming back to the holographic dark energy (HDE), the corresponding holographic cut-off $L_\mathrm{IR}$ is generally taken as particle horizon $L_\mathrm{p}$
or the future horizon $L_\mathrm{f}$.
However the fundamental form of the cut-off is still a debatable topic, and along this direction, it deserves mentioning that the most general holographic cut-off
was proposed in \cite{Nojiri:2005pu} where the $L_\mathrm{IR}$ is generalized to the form
$L_\mathrm{IR} = L_\mathrm{IR}\!\left(L_\mathrm{p}, \dot L_\mathrm{p}, \ddot L_\mathrm{p}, \cdots, L_\mathrm{f}, \dot L_\mathrm{f},
\ddot L_\mathrm{f}, \cdots, a, H, \dot H, \ddot H, \cdots\! \right)$ (where $a$ is the scale factor and $H$ is the Hubble parameter of the universe),
known as {\it generalized HDE}.
Based on this formalism, the known entropic dark energy models proposed so far are shown to be equivalent to holographic dark energy,
where the corresponding cut-off depends on either the particle horizon and its derivatives or the future horizon and its derivatives \cite{Nojiri:2021iko, Nojiri:2021jxf}.
Along this line, very recently the idea of generalized entropy (that can generalize a wide class of known entropies like the Bekenstein-Hawking entropy,
the Tsallis entropy, the Renyi entropy, the Sharma-Mittal entropy, the Kaniadakis entropy, etc.) has been proposed \cite{Nojiri:2022aof, Nojiri:2022dkr, Odintsov:2022qnn}
(one may also go through \cite{Odintsov:2023qfj} for a short review on generalized entropy), which also seems to have a {\it generalized holographic}
correspondence and can successfully unify viable slow-roll inflation with a viable dark energy era \cite{Nojiri:2022dkr}.
These reveal the rich cosmological implications of the generalized holographic formalism.

Based on the above discussions, such generalized form of $L_\mathrm{IR}$ immediately leads to the following questions:
\begin{itemize}
\item Does there exist suitable holographic cut-off(s) which can successfully drive constant roll inflation during the early universe?
If so, then what about the observable viability of such ``holographic constant roll inflation''?
\item Besides the mere inflation, does there exist any generalized $L_\mathrm{IR}$ that produces a unified cosmological scenario
from constant roll inflation to the dark energy era of the universe from a holographic point of view?
\end{itemize}

We will intend to address these questions in the present paper.
We should mention that the holographic constant roll inflation has been studied earlier in \cite{Mohammadi:2022vru}, however,
in a different context, the author of \cite{Mohammadi:2022vru} considered the parameter $c$
(appears in the holographic energy density $\rho_\mathrm{hol}$ as $\rho_\mathrm{hol} \propto c^2/L_\mathrm{IR}^2$) to be time-dependent.
On contrary, in our present analysis, we will consider the parameter $c$ to be a constant (more physical) and the infrared cut-off to be
of the generalized form to establish the holographic realization of constant roll inflation.
Besides inflation, we will also examine the holographic correspondence of the unification from constant roll inflation to the dark energy era.
These make the present work essentially different than the earlier one.

The paper is organized as follows: After going through some basics of constant roll inflation in scalar-tensor theory in Sec.~\ref{sec-CRI},
we will examine the holographic connection of constant roll inflation and the unification of constant roll inflation with the dark energy era
in Sec.~\ref{sec-hcr} and Sec.~\ref{sec-unification}, respectively.
In Sec.~\ref{sec-intermediate} between these two sections, we will consider a modified holographic constant roll inflation,
where the infrared cut-off is corrected by the ultraviolet one.
The paper ends with some conclusions in Sec.~\ref{sec-conclusions}.

\section{A brief on constant roll inflation}\label{sec-CRI}

In this section, we will revisit the main essence of constant roll inflation driven by a scalar field \cite{Motohashi:2014ppa},
where the action is given by,
\begin{align}
S = \int d^4x\sqrt{-g}\left[\frac{R}{2\kappa^2} - \frac{1}{2}g^{\mu\nu}\partial_{\mu}\phi\partial_{\nu}\phi - V(\phi)\right]\, .
\label{action-1}
\end{align}
Here $\kappa^2 = 8\pi G$ (with $G$ being Newton's gravitational constant) and $\phi$ is the scalar field under consideration with a potential $V(\phi)$.
The flat Friedmann-Lema\^{i}tre-Robertson-Walker metric fulfills our present purpose, in which case,
the Friedmann equations and the scalar field equation are given by,
\begin{align}
\frac{3}{\kappa^2} H^2=&\, \frac{1}{2}{\dot\phi}^2 + V(\phi)\, ,
\label{eq-1}\\
 -\frac{2}{\kappa^2}\dot{H}=&\, \dot{\phi}^2\, ,\label{eq-2}\\
0=&\, \ddot{\phi} + 3H\dot{\phi} + \frac{\partial V}{\partial\phi}\, ,
\label{eq-3}
\end{align}
respectively, where $H = \frac{d\ln{a}}{dt}$ is the Hubble parameter and $a(t)$ denotes the scale factor of the universe.
In the case of slow-roll inflation, the acceleration of the scalar field is ignored in the flat regime of $V(\phi)$,
owing to which, the first term in the right-hand side of Eq.~(\ref{eq-3}) is neglected with respect to the other terms.
However, the constant roll inflation deals in a different regime where $\ddot{\phi}$ is not negligible, rather $\ddot{\phi}/\left(H\dot{\phi}\right)$ is a constant.
In particular, the constant roll condition is given by,
\begin{align}
\ddot{\phi} = \beta H\dot{\phi}\, ,
\label{constant roll condition}
\end{align}
where $\beta$ is a constant (known as the constant roll parameter).
Clearly, for $\beta \approx 0$, the above condition can be thought of as equivalent to the standard slow-roll case.

Differentiating Eq.~(\ref{eq-2}) (with respect to $t$) and applying the constant roll condition, one obtains,
\begin{align}
\ddot{H} = 2\beta H\dot{H}\, ,
\label{eq-4}
\end{align}
on integrating which, gives a first-order differential equation of the Hubble parameter as,
\begin{align}
\dot{H} = \beta\left(H^2 - M^2\right)\, .
\label{eq-5}
\end{align}
Here $M$ is an integration constant considered to take positive values.
Moreover Eq.~(\ref{eq-2}) can be rewritten as,
\begin{align}
\dot{\phi} = -\left(\frac{2}{\kappa^2}\right)\frac{dH}{d\phi}\, ,
\label{eq-6}
\end{align}
by plugging which into Eq.~(\ref{eq-1}), yields the scalar field potential in terms of $H$ and $\frac{dH}{d\phi}$:
\begin{align}
V(\phi) = \frac{1}{\kappa^2}\left[3H^2 - \frac{2}{\kappa^2}\left(\frac{dH}{d\phi}\right)^2\right]\, .
\label{potential}
\end{align}
The above expression of $V(\phi)$ will be useful at some stage.
Having obtained the equations under constant roll conditions, we now move for the solutions of these field equations, and for this purpose,
we will consider two different cases depending on whether $\beta > 0$ or $\beta < 0$, respectively.

For $\beta > 0$, Eq.~(\ref{eq-5}) indicates that the Hubble parameter remains less than $M$ (as $\dot{H}$ should be negative),
and hence the solution of Eq.~(\ref{eq-5}) turns out to be,
\begin{align}
H(t) = -M\mathrm{tanh}\left(\beta Mt\right)\, .
\label{sol-1}
\end{align}
By using the solution of $H = H(t)$ into $\ddot{\phi} = \beta \dot{\phi}$ and integrating twice (with respect to the cosmic time),
one obtains the evolution of the scalar field as,
\begin{align}
\phi(t) = \frac{2}{\kappa}\sqrt{\frac{2}{\beta}}~\mathrm{tan}^{-1}\left(\e^{\beta Mt}\right)\, .
\label{sol-2}
\end{align}
The above $\phi = \phi(t)$ helps to eliminate $t$ from Eq.~(\ref{sol-1}) and results to the Hubble parameter in terms of $\phi$ as,
\begin{align}
H(\phi) = M\mathrm{cos}\left(\sqrt{\frac{\beta}{2}}~\kappa\phi\right)\, ,
\label{sol-3}
\end{align}
which along with Eq.~(\ref{potential}) immediately leads to the following form of $V(\phi)$ corresponding to the aforementioned solutions:
\begin{align}
V(\phi) = \frac{3M^2}{\kappa^2}\left[\left(\frac{3-\beta}{6}\right) + \left(\frac{3+\beta}{6}\right)\mathrm{cos}\left(\sqrt{2\beta}~\kappa\phi\right)\right]\, .
\label{sol-4}
\end{align}
Eq.~(\ref{sol-1}) depicts that the variable `$Mt$' ranges within $-\infty < Mt < 0$ to have a positive valued Hubble parameter (the positive Hubble parameter indicates an expanding universe where $t$ increases from $-\infty$ to zero). Thus during the early universe, i.e., during $Mt \ll -1$, the Hubble parameter from Eq.~(\ref{sol-1}) behaves as $H(t) \approx M$,
which results in a de-Sitter inflationary stage. Furthermore, the first slow-roll parameter is calculated as,
\begin{align}
 \epsilon_\mathrm{1} = -\frac{\dot{H}}{H^2} = \frac{\beta}{\mathrm{sinh}^2\left(\beta Mt\right)}\, .
 \label{SR-1}
\end{align}
This demonstrates that $\epsilon_\mathrm{1}$ is an increasing function with respect to the cosmic time, and therefore,
$\epsilon_\mathrm{1}$ must reach unity at some point in time depending on the value of $\beta$.
For example, if one takes $\beta = 0.02$, then $\epsilon_\mathrm{1}$ reaches to unity at $Mt = -7$.
The instance of $\epsilon_\mathrm{1} = 1$ indicates the end of inflation.
Consequently, the beginning of inflation (when the CMB scale $\sim 0.05\,\mathrm{Mpc}^{-1}$ crosses the horizon)
may be considered as 55 or 60 e-folds back from the instance of $\epsilon_\mathrm{1} = 1$ \cite{Motohashi:2014ppa}.
Thus as a whole, the model of action (\ref{action-1}) with $\beta > 0$ can drive a constant roll inflationary scenario,
which has an exit after 55 or 60 e-fold numbers. Here it deserves mentioning that the other case, i.e., $\beta < 0$, leads to a power law inflation during the early universe,
however, inflation is eternal and has no exit mechanism.
Keeping this in mind, we will consider the $\beta > 0$ case to establish the holographic equivalence of constant roll inflation.

\section{Holographic constant roll inflation}\label{sec-hcr}

In the holographic principle, the holographic energy density is proportional to the inverse squared infrared cutoff
$L_\mathrm{IR}$, which could be related to the causality given by the cosmological horizon,
\begin{equation}
\label{basic}
\rho_\mathrm{hol}=\frac{3c^2}{\kappa^2 {L_\mathrm{IR}}^2}\, .
\end{equation}
Here $c$ is a free parameter.
Identifying $\rho_\mathrm{hol}$ as the sole energy density, the Friedmann equation gives,
\begin{align}
\label{H2}
H=\frac{c}{L_\mathrm{IR}}\, .
\end{align}
Note that ${H}^{-1}$ corresponds to the radius of the cosmological horizon, which may correspond to the infrared cutoff scale, as expected.
As a candidate of the infrared cutoff $ L_\mathrm{IR}$, we may consider the particle horizon $L_\mathrm{p}$
or the future event horizon $L_\mathrm{f}$, which are given as follows,
\begin{equation}
\label{H3}
L_\mathrm{p}\equiv a \int_0^t\frac{dt}{a}\ ,\quad
L_\mathrm{f}\equiv a \int_t^\infty \frac{dt}{a}\, .
\end{equation}
Differentiating both sides of the above expressions lead to the Hubble parameter in terms of $L_\mathrm{p}$, $\dot{L}_\mathrm{p}$
or in terms of $L_\mathrm{f}$, $\dot{L}_\mathrm{f}$ as,
\begin{equation}
\label{HLL}
H \left( L_\mathrm{p} , \dot{L}_\mathrm{p} \right) = \frac{\dot{L}_\mathrm{p}}{L_\mathrm{p}} - \frac{1}{L_\mathrm{p}}\, , \quad
H(L_\mathrm{f} , \dot{L}_\mathrm{f}) = \frac{\dot{L}_\mathrm{f}}{L_\mathrm{f}} + \frac{1}{L_\mathrm{f}} \, .
\end{equation}
In \cite{Nojiri:2005pu}, a general form of the cutoff was proposed, which could be
a function of both $L_\mathrm{p}$ and $L_\mathrm{f}$ and their
derivatives, or additionally of the Hubble horizon and its derivatives as well as of the scale factor, namely,
\begin{align}
\label{generalLIR}
L_\mathrm{IR} = L_\mathrm{IR} \left(L_\mathrm{p}, \dot L_\mathrm{p}, \ddot L_\mathrm{p}, \cdots, L_\mathrm{f}, \dot L_\mathrm{f},
\ddot L_\mathrm{f}, \cdots, a, H, \dot H, \ddot H, \cdots \right)\, .
\end{align}
Based on this generalized formalism, we now establish that the constant roll inflation (discussed in the previous section)
has an equivalent holographic correspondence with a suitable cut-off.
In particular, by comparing Eq.~(\ref{eq-4}) with Eq.~(\ref{H2}), we may identify,
\begin{align}
\label{hcr1}
L_\mathrm{IR}=L^{(1)}_\mathrm{IR}\equiv \frac{2c \beta \dot H}{\ddot H} \, ,
\end{align}
or the comparison of Eq.~(\ref{eq-5}) with Eq.~(\ref{H2}) yields,
\begin{align}
\label{hcr2}
L_\mathrm{IR}=L^{(2)}_\mathrm{IR}\equiv \frac{c \beta H}{\dot H + \beta M^2} \, .
\end{align}
Moreover it may be noted that Eq.~(\ref{eq-5}) can be written as,
\begin{align}
\label{hcr-int1}
\frac{d}{dt} \left( H a^{-\beta} \right) = -\beta M^2 a^{-\beta} \, ,
\end{align}
on integrating which (with respect to the cosmic time), we obtain
\begin{align}
\label{hcr-int2}
H = -\beta M^2 a^\beta \int^t dt a^{-\beta} \, .
\end{align}
The above equation provides the general infrared cut-off, which resembles with
the particle horizon $L_\mathrm{p}$ or the future event horizon $L_\mathrm{f}$ as,
\begin{align}
\label{hcr4}
L^{(3)}_\mathrm{IR}\equiv -\frac{c}{\beta M^2a^\beta \int_0^t dt a^{-\beta}} \quad \mbox{or} \quad
L^{(4)}_\mathrm{IR}\equiv  \frac{c}{\beta M^2a^\beta \int_t^\infty dt a^{-\beta}}\, ,
\end{align}
respectively. Therefore in the present case, the generalized holographic cut-off can be expressed by either of the following ways:
\begin{align}
L_\mathrm{IR} \equiv
\begin{cases}
L^{(1)}_\mathrm{IR}=2c \beta \dot H/\ddot H~~~~~~~~~~~~~~~~~~~~~~\mathrm{or}~~~~~~~~~~~~L^{(2)}_\mathrm{IR}=c \beta H/\left(\dot H + \beta M^2\right)~~\mathrm{or} & \\
L^{(3)}_\mathrm{IR}=-c/\left(\beta M^2a^\beta \int_0^t dt a^{-\beta}\right)~~~~~~~~\mathrm{or}~~~~~~~~L^{(4)}_\mathrm{IR}=c/\left(\beta M^2a^\beta \int_t^\infty dt a^{-\beta}\right)~~.
\end{cases}
\label{hcr-rev2}
\end{align}
Furthermore, due to Eq.~(\ref{HLL}), the above forms of $L^{(1)}_\mathrm{IR}$ and $L^{(2)}_\mathrm{IR}$ can also be expressed either in terms of particle horizon
and its derivatives or in terms of the future horizon and its derivatives (see the Appendix, Sec.~\ref{appendix}).

The holographic Friedmann equation $H = \frac{c}{L_\mathrm{IR}}$ with the cut-off given by $L^{(1)}_\mathrm{IR}$
reproduces the cosmological field Eq.~(\ref{eq-4}), or similarly, the cut-off $L^{(2)}_\mathrm{IR}$, $L^{(3)}_\mathrm{IR}$, or $L^{(4)}_\mathrm{IR}$ results to Eq.~(\ref{eq-5}).
Here we need to recall that Eq.~(\ref{eq-4}) or Eq.~(\ref{eq-5}) can lead to constant roll inflation as discussed in Sec.~\ref{sec-CRI}.
Thus we may argue that the infrared cut-offs of the form $L^{(1)}_\mathrm{IR}$ as well as of the form $L^{(2)}_\mathrm{IR}$,
$L^{(3)}_\mathrm{IR}$, or $L^{(4)}_\mathrm{IR}$ are equally able to drive constant roll inflation, which we may call ``holographic constant roll inflation''.
This indeed establishes the holographic equivalence of the constant roll inflation in the present context. We may also
write the general holographic cut-off corresponds to the constant roll inflation as follows:
\begin{align}
\label{hcr5}
\frac{1}{L_\mathrm{IR}} = \frac{c^{(1)}}{L^{(1)}_\mathrm{IR}} + \frac{c^{(2)}}{L^{(2)}_\mathrm{IR}} + \frac{c^{(3)}}{L^{(3)}_\mathrm{IR}}
+ \frac{c^{(4)}}{L^{(4)}_\mathrm{IR}}\, ,
\end{align}
where the coefficients satisfy $c^{(1)} + c^{(2)} + c^{(3)} + c^{(4)} = 1$.
The effective equation of state (EoS) parameter corresponding to the above $L_\mathrm{IR}$ is given by,
\begin{align}
\omega_\mathrm{hol} = -1 + \left(\frac{2}{3c}\right)\frac{dL_\mathrm{IR}}{dt}\, .
\label{eos-hol}
\end{align}
Having the holographic model of Eq.~(\ref{hcr5}) in hand, we now determine the evolution of the $L_\mathrm{IR}$.
With the explicit forms of $L^{(i)}_\mathrm{IR}$ ($i = 1,2,3,4$), Eq.~(\ref{hcr5}) turns out to be,
\begin{align}
\frac{1}{L_\mathrm{IR}} = c^{(1)}\left(\frac{\ddot{H}}{2c\beta\dot{H}}\right) + c^{(2)}\left(\frac{\dot{H}
+ \beta M^2}{c\beta H}\right) - c^{(3)}\left(\frac{\beta M^2a^{\beta}\int_{0}^{t}a^{-\beta}dt}{c}\right)
+ c^{(4)}\left(\frac{\beta M^2a^{\beta}\int_{t}^{\infty}a^{-\beta}dt}{c}\right)\, .
\label{hcr6}
\end{align}
Using the holographic Friedmann equation $H = \frac{c}{L_\mathrm{IR}}$ along with the aforementioned constraint relation $c^{(1)} + c^{(2)} + c^{(3)} + c^{(4)} = 1$,
the above equation yields the following solution of the $L_\mathrm{IR}$:
\begin{align}
\label{hcr7}
\frac{c}{L_\mathrm{IR}} = -M \tanh \left( \beta M t \right) \, .
\end{align}
Here it deserves mentioning that the above solution of $L_\mathrm{IR}$ satisfies
\begin{align}
\frac{\ddot{L}_\mathrm{IR} - 2\left(\dot{L}_\mathrm{IR}\right)^2/L_\mathrm{IR}}{2c\left(\dot{L}_\mathrm{IR}/L_\mathrm{IR}\right)} = \beta~(\mbox{constant})\, ,
\label{hcr8}
\end{align}
which is the constant roll condition in a holographic scenario.
Therefore the solution of the holographic cut-off obtained in Eq.~(\ref{hcr7}) indeed points to constant roll inflation.
This solution of $L_\mathrm{IR}$, in turn, helps to calculate the observable quantities, by which,
we can investigate the viability of the holographic inflationary scenario.
The slow-roll quantities in holographic inflation are given by,
\begin{align}
\epsilon_\mathrm{1}=&\, -\frac{\dot{H}}{H^2} = \frac{1}{c}\frac{dL_\mathrm{IR}}{dt}\, ,\nonumber\\
\epsilon_\mathrm{n+1}=&\, \frac{\dot{\epsilon}_\mathrm{n}}{H\epsilon_\mathrm{n}} = \frac{\dot{\epsilon}_\mathrm{n}}{\epsilon_\mathrm{n}\left(c/L_\mathrm{IR}\right)}\, ,
\label{hcr9}
\end{align}
with $n \geq 1$. Consequently the scalar spectral index ($n_s$) and the tensor-to-scalar ratio ($r$) come as,
\begin{align}
n_s = \left. \left[1 - 2\epsilon_\mathrm{1} - 2\epsilon_\mathrm{2}\right]\right|_\mathrm{h.c.} \quad \mbox{and} \quad
\left. r = 16\epsilon_\mathrm{1}\right|_\mathrm{h.c.}\, ,
\label{hcr10}
\end{align}
respectively, where the suffix `$\mathrm{h.c.}$' symbolizes the horizon crossing instant of the CMB scale mode in which we are interested.
Here we would like to mention that the above expressions of $n_s$ and $r$ are valid for slow-roll inflation.
However, we will show that the present constant roll inflationary model obtains compatibility with the Planck data for $\beta \ll 1$;
thus we safely work with Eq.~(\ref{hcr10}) even in the present context.
The $L_\mathrm{IR}$ immediately leads to $\epsilon_\mathrm{1}$ and $\epsilon_\mathrm{2}$ as follows:
\begin{align}
\epsilon_\mathrm{1}=&\, \frac{\beta}{\mathrm{sinh}^2\left(\beta Mt\right)} = \beta\left(\frac{M^2{L_\mathrm{IR}}^2}{c^2} - 1\right)\, ,\nonumber\\
\epsilon_\mathrm{2}=&\, \frac{2\beta}{\mathrm{tanh}^2\left(\beta Mt\right)} = \frac{2\beta M^2{L_\mathrm{IR}}^2}{c^2}\, .
\label{hcr11}
\end{align}
Plugging the above expressions of the slow-roll parameters into Eq.~(\ref{hcr10}) and after a little bit of simplification,
we obtain the final forms of $n_s$ and $r$ as:
\begin{align}
n_s=&\, \left. 1 - 2\beta\left(\frac{3M^2{L_\mathrm{IR}}^2}{c^2} - 1\right)\right|_\mathrm{h.c.}\, ,\nonumber\\
r=&\, \left. 16\beta\left(\frac{M^2{L_\mathrm{IR}}^2}{c^2} - 1\right)\right|_\mathrm{h.c.}\, .
\label{hcr12}
\end{align}
It should be noticed that $n_s$ and $r$ depend on the dimensionless parameters $\frac{ML_\mathrm{IR}\left(t_\mathrm{h} \right)}{c}$
and $\beta$ (where $L_\mathrm{IR}\left(t_\mathrm{h} \right)$ is the infrared cut-off at the time of horizon crossing of the CMB scale mode
with $t_\mathrm{h}$ being the horizon crossing instance).
We can now directly confront the spectral index and the tensor-to-scalar ratio with the Planck 2018 results \cite{Planck:2018jri},
which constrain the observational indices to be:
\begin{align}
n_s = 0.9649 \pm 0.0042 \quad \mbox{and} \quad r < 0.064\, .
\label{hcr13}
\end{align}

 \begin{figure}[!h]
\begin{center}
 \centering
 \includegraphics[width=3.5in,height=2.5in]{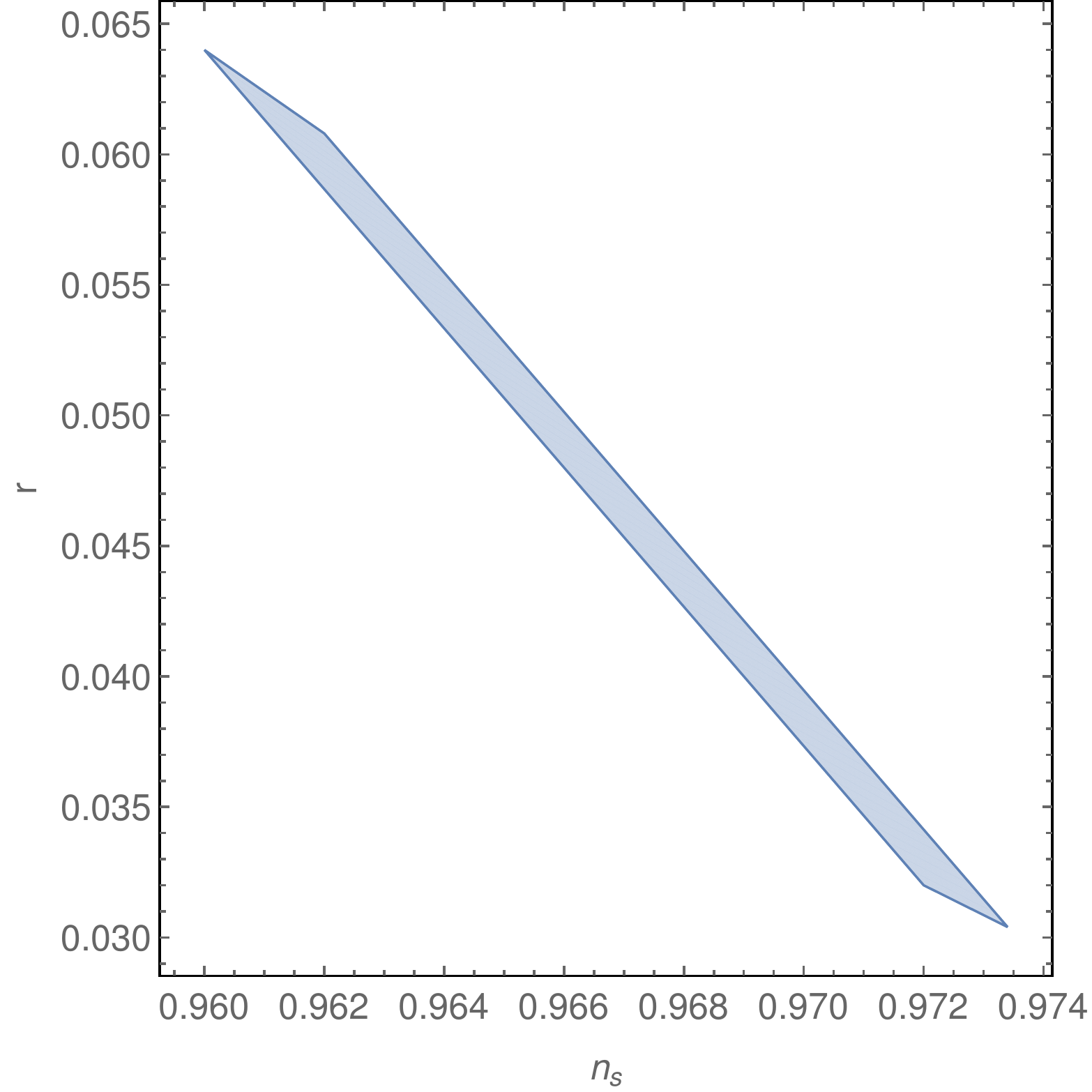}
 \caption{Parametric plot of $n_s$ vs. $r$ for the following parameter ranges: $\sqrt{\frac{3}{2}} < \frac{ML_\mathrm{IR}(t_\mathrm{h})}{c} < \sqrt{2}$ and $0.0038 < \beta < 0.004$.}
 \label{plot observable}
\end{center}
\end{figure}

For the holographic model at hand, the $n_s$ and $r$ prove to be simultaneously compatible with the Planck constraints
for the following narrow ranges of the parameters:
\begin{align}
\sqrt{\frac{3}{2}} < \frac{ML_\mathrm{IR}\left(t_\mathrm{h} \right)}{c} < \sqrt{2} \quad \mbox{and} \quad ~0.0038 < \beta < 0.004\, .
 \label{hcr-final1}
\end{align}
This is depicted in FIG.~\ref{plot observable}.
Thus as a whole, the holographic model of Eq.~(\ref{hcr6}) triggers constant roll inflation which is indeed compatible with the Planck data
provided the holographic cut-off at horizon crossing (i.e., $L_\mathrm{IR}\left(t_\mathrm{h}\right)$) and the constant roll parameter $\beta$ lie
within the above-mentioned range.

\section{More on holographic constant roll inflation}\label{sec-intermediate}

In this section, following \cite{Nojiri:2019kkp}, we will consider an ultraviolet correction over the infrared one,
and thus the holographic cut-off takes the following form,
\begin{align}
\label{H8c}
L \equiv \sqrt{ {L_\mathrm{IR}}^2 + \frac{1}{{\Lambda_\mathrm{UV}}^2}}\, .
\end{align}
During the inflationary era, i.e., at high energy scales, the quantum gravity effects may become important,
and hence the infrared cut-off acquires a correction by the ultraviolet one.
In such modified holographic scenario, the comparison of $H = \frac{c}{L}$ with Eq.~(\ref{eq-4}) or with Eq.~(\ref{eq-5})
yields the infrared cut-off(s) corresponding to the constant roll inflation as follows,
\begin{equation}
L_\mathrm{IR}=L^{(1)}_\mathrm{IR}\equiv \sqrt{\left(\frac{2c \beta \dot H}{\ddot H}\right)^2 - \frac{1}{{\Lambda_\mathrm{UV}}^2}} \, ,
\label{corr-1}
\end{equation}
or,
\begin{align}
L_\mathrm{IR}=L^{(2)}_\mathrm{IR}\equiv \sqrt{\left(\frac{c \beta H}{\dot H + \beta M^2}\right)^2 - \frac{1}{{\Lambda_\mathrm{UV}}^2}} \, ,
\label{corr-2}
\end{align}
respectively.
Due to the holographic Friedmann equation $H = \frac{c}{L}$, the cut-off $L^{(1)}_\mathrm{IR}$ clearly reproduces Eq.~(\ref{eq-4}), or similarly,
the $L^{(2)}_\mathrm{IR}$ reproduces Eq.~(\ref{eq-5}).
Owing to this, we may argue that $L^{(1)}_\mathrm{IR}$ or $L^{(2)}_\mathrm{IR}$, in the presence of ultraviolet cut-off, can drive constant roll inflation.
We can also determine the evolution of such infrared cut-off(s), and is given by,
\begin{align}
L^{(1)}_\mathrm{IR} = L^{(2)}_\mathrm{IR} = \sqrt{\frac{c^2}{M^2\mathrm{tanh}^2\left(\beta Mt\right)} - \frac{1}{{\Lambda_\mathrm{UV}}^2}}\, .
\label{corr-3}
\end{align}
Consequently, the scalar spectral index and the tensor-to-scalar ratio in the present context turn out to be,
\begin{align}
n_s=&\, \left. 1 - 2\beta\left\{\frac{3M^2}{c^2}\left({L_\mathrm{IR}}^2 + \frac{1}{{\Lambda_\mathrm{UV}}^2}\right) - 1\right\}\right|_\mathrm{h.c.}\, ,\nonumber\\
r=&\, \left. 16\beta\left\{\frac{M^2}{c^2}\left({L_\mathrm{IR}}^2 + \frac{1}{{\Lambda_\mathrm{UV}}^2}\right) - 1\right\}\right|_\mathrm{h.c.}\, .
\label{corr-4}
\end{align}
The above theoretical expectations of $n_s$ and $r$ in such a modified holographic scenario turn out to be simultaneously consistent
with the Planck 2018 constraints (see Eq.~(\ref{hcr13})) provided $L_\mathrm{IR}\left(t_\mathrm{h} \right)$ (i.e., the infrared cut-off at the instant of horizon crossing)
and $\beta$ lie within the following ranges:
\begin{align}
\sqrt{\frac{3c^2}{2M^2} - \frac{1}{{\Lambda_\mathrm{UV}}^2}} < L_\mathrm{IR}\left(t_\mathrm{h} \right) < \sqrt{\frac{2c^2}{M^2} - \frac{1}{{\Lambda_\mathrm{UV}}^2}}
\quad \mbox{and} \quad 0.0038 < \beta < 0.004\, .
\label{corr-5}
\end{align}
It is clear by comparing Eq.~(\ref{corr-5}) with Eq.~(\ref{hcr-final1}) that in presence of the ultraviolet correction,
the constraint on $L_\mathrm{IR}\left(t_\mathrm{h} \right)$ becomes different compared to the previous case where there is no ultraviolet correction,
and moreover, for $\Lambda_\mathrm{UV} \rightarrow \infty$, the above constraint on $L_\mathrm{IR}\left(t_\mathrm{h} \right)$ resembles with that of in Eq.~(\ref{hcr-final1}).
Therefore in the present modified holographic scenario, the appearance of $\Lambda_\mathrm{UV}$ may give some extra freedom to some extent (one may go through \cite{Nojiri:2019kkp}).

\section{Unification from constant roll inflation to dark energy: Holographic view}\label{sec-unification}

The constant roll inflation in (\ref{sol-1}) well describes the inflation in the early epoch corresponding to $t<0$,
however, the Hubble rate $H$ during $t > 0$ becomes negative and therefore it does not describe the expansion of the universe.
In the spirit of this, we may consider the following simple modification of the Hubble parameter:
\begin{align}
\label{hcr19}
H(t) = - M \tanh \left( \gamma M t \right) + M + h_0 \, ,
\end{align}
for which, the associated scale factor of the universe is given by: $a \propto \cosh^{-\frac{1}{\gamma}} \left( \gamma M t \right) \e^{\left( M + h_0 \right)t}$
where $h_0$, $M$, and $\gamma$ are positive parameters (it may be mentioned that both $h_0$ and $M$ have mass dimension $[+1]$
while $\gamma$ is dimensionless). Here the constant roll parameter is replaced by $\gamma$ (in place of $\beta$) to differentiate the present case from the previous ones. Eq.~(\ref{hcr19}) clearly depicts that for $t \rightarrow - \infty$, $H$ becomes a constant, in particular, $H \rightarrow 2M + h_0$,
which results in a de-Sitter inflationary stage.
On the other hand, when $H \rightarrow h_0$ becomes a constant again at $t \rightarrow +\infty$, which may correspond to the dark energy.
Moreover the Hubble parameter in Eq.~(\ref{hcr19}) satisfies,
\begin{align}
\frac{\ddot{H}}{2H\dot{H}} = \frac{\gamma M\mathrm{tanh}\left(\gamma Mt\right)}{M\mathrm{tanh}\left(\gamma Mt\right) - M - h_0}\, .
\end{align}
Thus the quantity $\frac{\ddot{H}}{2H\dot{H}}$ tends to a positive constant value at the large negative time, which in turn ensures that the inflation is a constant roll in nature.
Therefore by the modification in (\ref{hcr19}), we can describe both the constant roll inflation in the early universe and dark energy in the late universe in a unified manner.
Because Eq.~(\ref{hcr19}) is a shift by a constant $M + h_0$ from (\ref{sol-1}), Eq.~(\ref{eq-4}) and Eq.~(\ref{eq-5}) are modified as follows,
\begin{align}
\label{hcr21}
\ddot H - 2\gamma \left( H - M-h_0 \right)\dot H=&\, 0\, ,\nonumber\\
\dot H - \gamma \left( H - M - h_0 \right)^2=&\, -\gamma M^2\, .
\end{align}
Such a unified picture of constant roll inflation and dark energy may be described by some suitable higher curvature
like $f(R)$ gravity theory, for instance, see \cite{Odintsov:2017hbk} where $f(R)$ initially takes an $R^2$ corrected logarithmic form
which drives constant roll inflation during the early phase, while at a late time, the $f(R)$ is considered to be more-or-less
of exponential form that ensures a viable dark energy era of the universe. In addition the standard cosmological evolution in-between the inflation and the dark energy era, and also the transition from the standard cosmology to the late time acceleration, have been successfully demonstrated in \cite{Odintsov:2017hbk}.
Or even a scalar-tensor theory may be also suitable to trigger $H(t)$ of Eq.~(\ref{hcr19}) -- where the scalar field initially rolls
at a constant rate to give the constant roll inflation and at a late time, the scalar field energy density acts
like a bare cosmological constant which in turn leads to a late dark energy era.
However, in the present context, our main concern is to establish the holographic equivalence of such a unified cosmological description,
rather than finding suitably modified gravity theory(ies) corresponding to this evolution.
For this purpose, let us individually compare the first and second equations of Eq.~(\ref{hcr21}) with Eq.~(\ref{H2}), which immediately identify,
\begin{align}
\label{hcr22}
L_\mathrm{IR}=&\, {\tilde L}^{(1)}_\mathrm{IR}\equiv \frac{2c\gamma}{\frac{\ddot H}{\dot H}+ 2\gamma\left( M + h_0 \right)} \, , \nonumber \\
L_\mathrm{IR}=&\, {\tilde L}^{(2)}_\mathrm{IR}\equiv \frac{c \gamma}{\frac{\dot H + \gamma M^2}{H - M - h_0} + \gamma \left(M + h_0 \right)} \, ,
\end{align}
respectively.
On the other hand, we may note that the second equation in Eq.~(\ref{hcr21}) can be rewritten as
\begin{align}
\label{hcr25}
\frac{1}{H}=\gamma~a^{-\beta} \int^t dt \left( \frac{2 \left(M+h_0\right)}{H} - \frac{2 h_0 M + h_0^2}{H^2} + 1 \right) a^\beta\, ,
\end{align}
on comparing which with Eq.~(\ref{H2}) identifies the following cut-offs in more-or-less similar fashion of the particle horizon $L_\mathrm{p}$ or the future event horizon $L_\mathrm{f}$:
\begin{align}
\label{hcr26}
{\tilde L}^{(3)}_\mathrm{IR}\equiv&\, \gamma c a^{-\beta} \int_0^t dt \left( \frac{2 \left(M+h_0\right)}{H} - \frac{2 h_0 M + h_0^2}{H^2} + 1 \right) a^\beta \, , \nonumber \\
{\tilde L}^{(4)}_\mathrm{IR}\equiv&\, - \gamma c a^{-\beta} \int_t^\infty dt \left( \frac{2 \left(M+h_0\right)}{H} - \frac{2 h_0 M + h_0^2}{H^2} + 1 \right) a^\beta \, ,
\end{align}
respectively. Therefore the generalized holographic cut-off, that results to an unification of constant roll inflation and dark energy era, can be expressed by either of the following ways:
\begin{align}
\tilde{L}_\mathrm{IR} \equiv
\begin{cases}
\tilde{L}^{(1)}_\mathrm{IR}= \mathrm{given~in~the~first~expression~of~Eq.(\ref{hcr22})}~,~~\mathrm{or} & \\
\tilde{L}^{(2)}_\mathrm{IR}= \mathrm{given~in~the~second~expression~of~Eq.(\ref{hcr22})}~,~~\mathrm{or} & \\
\tilde{L}^{(3)}_\mathrm{IR}= \mathrm{given~in~the~first~expression~of~Eq.(\ref{hcr26})}~,~~\mathrm{or} & \\
\tilde{L}^{(4)}_\mathrm{IR}= \mathrm{given~in~the~second~expression~of~Eq.(\ref{hcr26})}~~.
\end{cases}
\label{hcr-rev3}
\end{align}
The holographic Friedmann equation $H = \frac{c}{L_\mathrm{IR}}$ with cut-off identified as ${\tilde L}^{(1)}_\mathrm{IR}$
reproduces the first equation of Eq.~(\ref{hcr21}), or similarly, the holographic cut-offs like ${\tilde L}^{(2)}_\mathrm{IR}$, ${\tilde L}^{(3)}_\mathrm{IR}$,
or ${\tilde L}^{(4)}_\mathrm{IR}$ reproduce the second equation of Eq.~(\ref{hcr21}). This reveals that ${\tilde L}^{(1)}_\mathrm{IR}$,
${\tilde L}^{(2)}_\mathrm{IR}$, ${\tilde L}^{(3)}_\mathrm{IR}$, or ${\tilde L}^{(4)}_\mathrm{IR}$ is equally capable to demonstrate
the unified evolution of constant roll inflation and the dark energy era of the universe.
We may also express the general holographic cut-off corresponding to the such unified description as follows:
\begin{align}
\frac{1}{{\tilde L}_\mathrm{IR}} = \frac{c^{(1)}}{{\tilde L}^{(1)}_\mathrm{IR}} + \frac{c^{(2)}}{{\tilde L}^{(2)}_\mathrm{IR}}
+ \frac{c^{(3)}}{{\tilde L}^{(3)}_\mathrm{IR}} + \frac{c^{(4)}}{{\tilde L}^{(4)}_\mathrm{IR}}\, ,
\label{hcr27}
\end{align}
with $c^{(1)} + c^{(2)} + c^{(3)} + c^{(4)} = 1$. Due to $H = \frac{c}{{\tilde L}_\mathrm{IR}}$ along with the explicit forms of ${\tilde L}^{(i)}_\mathrm{IR}$ ($i = 1,2,3,4$),
we determine the evolution of the cut-off as follows:
\begin{align}
 \frac{c}{{\tilde L}_\mathrm{IR}} = - M \tanh \left( \gamma M t \right) + M + h_0\, .
 \label{hcr28}
\end{align}
${\tilde L}_\mathrm{IR}$ becomes constant at asymptotic values of the cosmic time which, due to $H = \frac{c}{{\tilde L}_\mathrm{IR}}$,
points to the unification of two accelerating stages of the universe from the holographic point of view. In particular,
for large negative and for large positive time, the ${\tilde L}_\mathrm{IR}$ becomes
${\tilde L}_\mathrm{IR} = \frac{c}{2M + h_0}$ and ${\tilde L}_\mathrm{IR} = \frac{c}{h_0}$, respectively, which depict holographic inflation
during the early stage and holographic dark energy during the late time of the universe.
Thus if we consider $h_0 \ll M$, then $M$ specifies the inflationary energy scale while $h_0$ is the Hubble scale during the dark energy stage.
Moreover the ${\tilde L}_\mathrm{IR}$ of Eq.~(\ref{hcr28}) obeys,
\begin{align}
\frac{\ddot{{\tilde L}}_\mathrm{IR} - 2 \left(\dot{{\tilde L}}_\mathrm{IR}\right)^2/{\tilde L}_\mathrm{IR}}{2c\left(\dot{{\tilde L}}_\mathrm{IR}/{\tilde L}_\mathrm{IR}\right)}
\rightarrow \mbox{constant}  \quad \mbox{at} \quad |Mt| \gg 1\, .
\label{hcr29}
\end{align}
This ensures that the holographic inflation occurring during the early universe is indeed a ``constant roll'' in nature.
Thus we may argue that the holographic model of Eq.~(\ref{hcr27}) can describe constant roll inflation and dark energy of the universe in a unified way.
Having obtained such a cut-off, we now examine the observable viability of the model with respect to the recent Planck data.
The first and second slow-roll parameters (defined in Eq.~(\ref{hcr9})) in this context turn out to be,
\begin{align}
\epsilon_\mathrm{1} = \frac{\gamma\left\{M^2 - \left(\frac{c}{{\tilde L}_\mathrm{IR}} - M - h_0\right)^2\right\}}{\left(c^2/\left({\tilde L}_\mathrm{IR}\right)^2\right)}
\quad \mbox{and} \quad \epsilon_\mathrm{2} = \frac{2\gamma\left\{M^2 + \left(M + h_0\right)
\left(\frac{c}{{\tilde L}_\mathrm{IR}} - M - h_0\right)\right\}}{\left(c^2/\left({\tilde L}_\mathrm{IR}\right)^2\right)}\, ,
\label{hcr30}
\end{align}
respectively, where we use Eq.~(\ref{hcr28}).
Consequently, the scalar spectral index and the tensor-to-scalar ratio take the following forms,
\begin{align}
n_s=&\, \left. 1-2\epsilon_\mathrm{1} - 2\epsilon_\mathrm{2}\right|_\mathrm{h.c.}
= \left. 1 - 2\gamma\left\{\frac{3M^2\left({\tilde L}_\mathrm{IR}\right)^2}{c^2} - \left(1 - \left(M + h_0\right)\frac{{\tilde L}_\mathrm{IR}}{c}\right)
\left(1 - 3\left(M + h_0\right)\frac{{\tilde L}_\mathrm{IR}}{c}\right)\right\}\right|_\mathrm{h.c.}\, ,\nonumber\\
r=&\, \left. 16\epsilon_\mathrm{1}\right|_\mathrm{h.c.}
= \left. 16\gamma\left\{\frac{M^2\left({\tilde L}_\mathrm{IR}\right)^2}{c^2} - \left(1 - \left(M + h_0\right)\frac{{\tilde L}_\mathrm{IR}}{c}\right)^2\right\} \right|_\mathrm{h.c.}\, ,
\label{hcr31}
\end{align}
respectively, with $t_\mathrm{h}$ being the horizon crossing instant of the CMB scale mode on which we are interested to evaluate the observable indices.
It is evident that both $n_s$ and $r$ in the present holographic scenario depend on $\frac{M{\tilde L}_\mathrm{IR}\left(t_\mathrm{h} \right)}{c}$,
$\frac{h_0{\tilde L}_\mathrm{IR}\left(t_\mathrm{h} \right)}{c}$ and $\gamma$.
It turns out that $n_s$ and $r$ are simultaneously compatible with the Planck 2018 data \cite{Planck:2018jri} provided $h_0 \ll M$
and the other parameters lie within the following ranges:
\begin{align}
0.80 < \frac{M{\tilde L}_\mathrm{IR}\left(t_\mathrm{h} \right)}{c} < 0.89 \quad \mbox{and} \quad 0.010 < \gamma < 0.011 \, .
\label{hcr32}
\end{align}

 \begin{figure}[!h]
\begin{center}
 \centering
 \includegraphics[width=3.5in,height=2.5in]{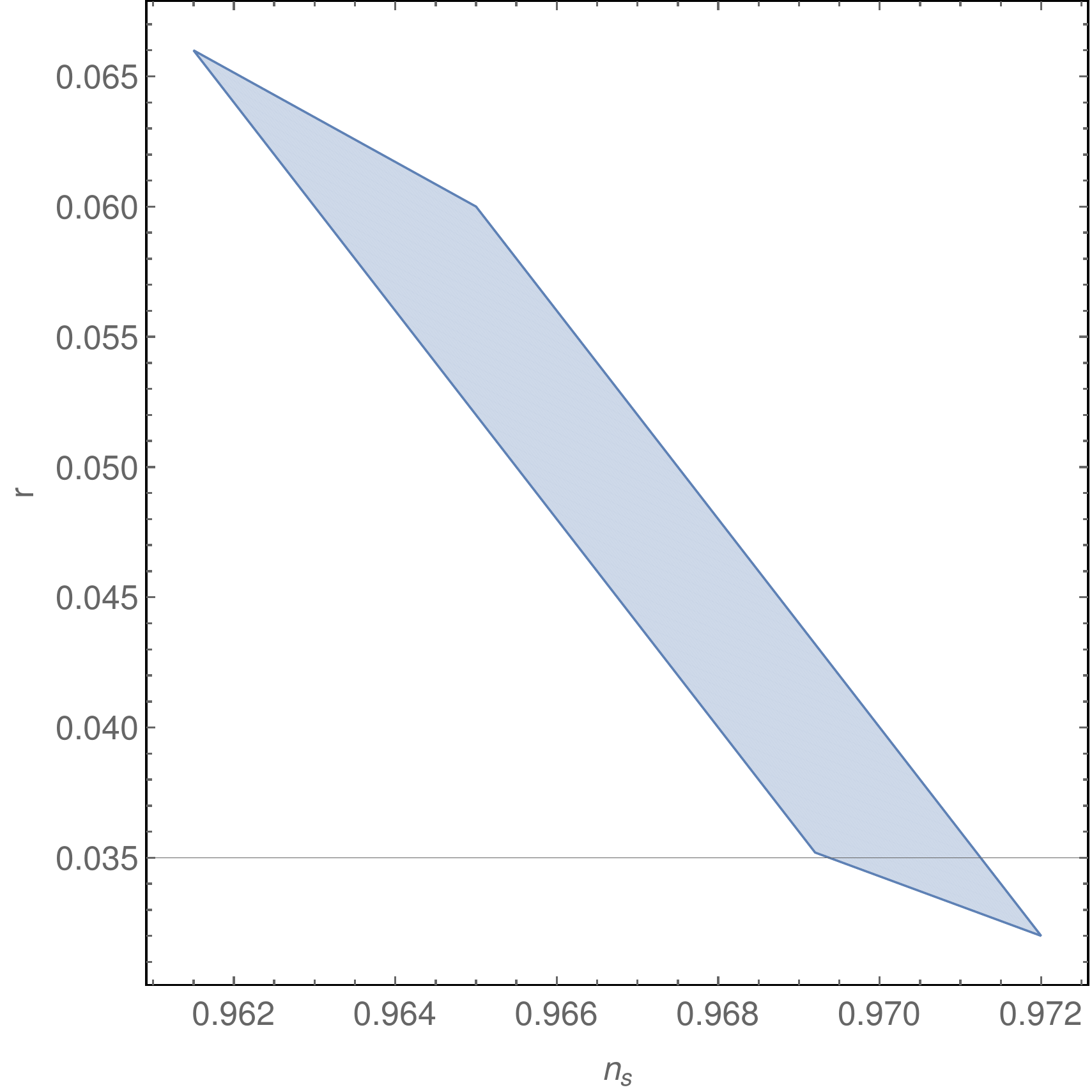}
 \caption{Parametric plot of $n_s$ vs. $r$ for the following parameter ranges: $0.80 < \frac{M{\tilde L}_\mathrm{IR}(t_\mathrm{h})}{c} < 0.89$, $0.010 < \gamma < 0.011$ and $h_0 \ll M$.}
 \label{plot observable-2}
\end{center}
\end{figure}

This is revealed in Fig.~\ref{plot observable-2}.
On the other hand, the effective EoS parameter corresponding to the cut-off in Eq.~(\ref{hcr27}) (or in Eq.~(\ref{hcr28})) comes as,
\begin{align}
\widetilde{\omega} = -1 + \left(\frac{2}{3c}\right)\frac{d{\tilde L}_\mathrm{IR}}{dt}
= -1 + \frac{2\gamma M^2\mathrm{sech}^2\left(\gamma Mt\right)}{3\left[-M\mathrm{tanh}\left(\gamma Mt\right) + M + h_0\right]^2}\, ,
\label{hcr33}
\end{align}
which tends to $\widetilde{\omega} \rightarrow -1$ at $Mt \gg 1$ i.e., during the dark energy era.
This is, however, expected as the cut-off in Eq.~(\ref{hcr28}) tends to a constant value, in particular, ${\tilde L}_\mathrm{IR} \approx \frac{c}{h_0}$,
during the large positive time.
Thus we take the parameter $h_0$ to be equal to the present Hubble parameter of the universe, i.e., $h_0 = H_0 = 10^{-33}\mathrm{eV}$ (where $H_0$
is the current Hubble parameter), which in turn makes the holographic cut-off at present time as ${\tilde L}^{(0)}_\mathrm{IR} = c\times10^{33}\mathrm{eV}$.
As a whole, the inflationary and the dark energy constraints in the present holographic scenario are given by,
\begin{align}
0.80 < \frac{M{\tilde L}_\mathrm{IR}\left(t_\mathrm{h} \right)}{c} < 0.89\, , \quad 0.010 < \gamma < 0.011
\quad \mbox{and} \quad h_0 \approx 10^{-34}\mathrm{eV} \, .
\label{hcr34}
\end{align}

\begin{figure}[!h]
\begin{center}
 \centering
 \includegraphics[width=3.5in,height=2.5in]{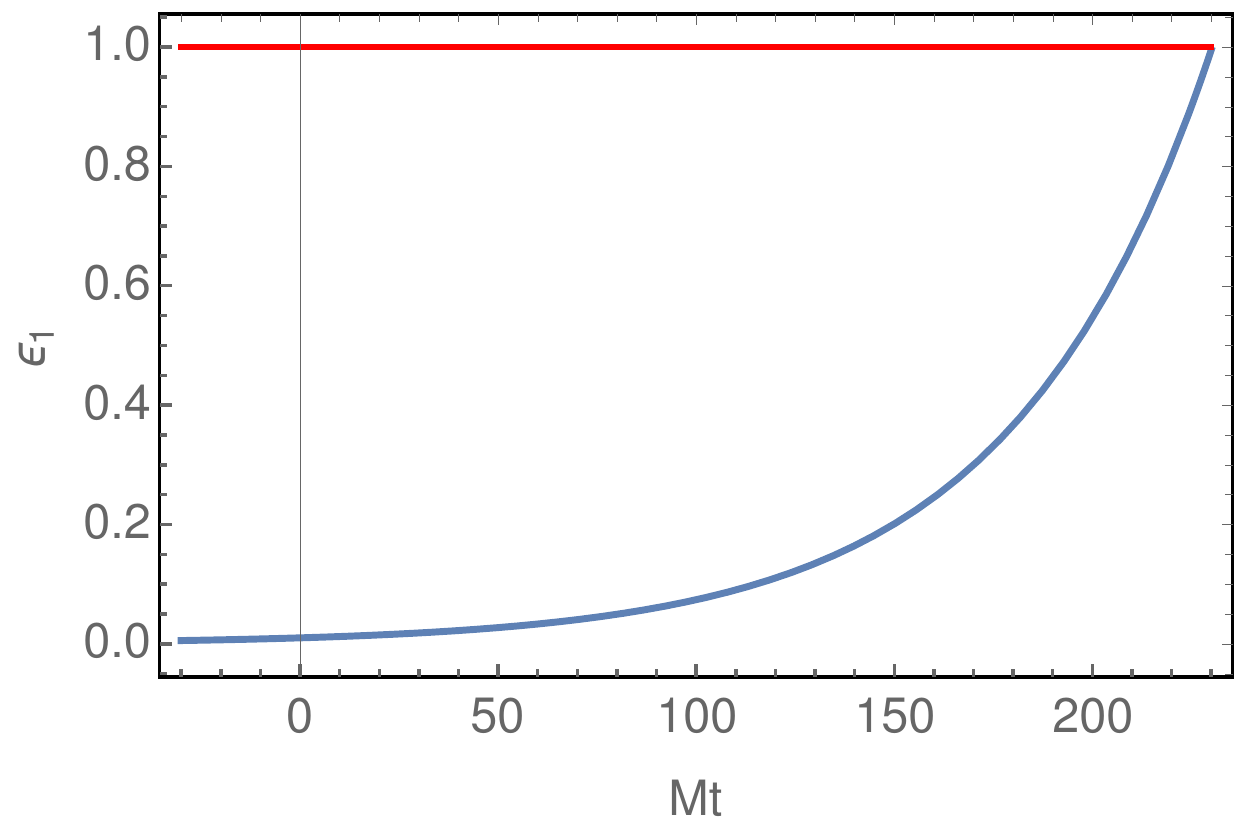}
 \caption{$\epsilon_\mathrm{1}$ vs. $Mt$ for $\gamma = 0.01$ and $h_0 = 10^{-34}\mathrm{eV} = 10^{-43}\mathrm{GeV}$ and $M = 10^{14}\mathrm{GeV}$. Such parametric ranges are consistent as of Eq.~(\ref{hcr34}).}
 \label{plot-new1}
\end{center}
\end{figure}

Having demonstrated the inflation and the dark energy era, we now like to describe the intermediate evolution of the universe (in-between the end of inflation and the dark energy era) in order to have an unified cosmology in the present holographic scenario. Regarding the end of inflation, we use the explicit evolution of $\tilde{L}_\mathrm{IR}$ from Eq.~(\ref{hcr28}) to Eq.~(\ref{hcr30}), and after a little bit of simplification, we obtain
\begin{eqnarray}
 \epsilon_\mathrm{1} = \frac{\gamma M^2~\mathrm{sech}^2\left(\gamma Mt\right)}{\left(-M\mathrm{tanh}(\gamma Mt) + M + h_0\right)^2}~~,
 \label{hcr35}
\end{eqnarray}
which has the behaviour as shown in Fig.[\ref{plot-new1}]. The figure clearly depicts that $\epsilon_\mathrm{1}$ is a monotonic increasing function with time and reaches to unity nearly at $Mt_\mathrm{f} = 230$ which, in turn, indicates the end of inflation ($t_\mathrm{f}$ represents the cosmic time when the inflation ends).

Therefore the beginning of inflation (when the CMB scale $\sim 0.05\mathrm{Mpc}^{-1}$ crosses the horizon) may be considered 50 to 65 e-folds back from the instance of $\epsilon_\mathrm{1} = 1$ (or equivalently $Mt = 230$). The e-fold number is given by $N = \int_{t_\mathrm{i}}^{t}~H dt$ which, due to the Hubble parameter of Eq.~(\ref{hcr19}), becomes
\begin{eqnarray}
 N = M\left(t - t_\mathrm{i}\right) + h_0\left(t - t_\mathrm{i}\right) - \frac{1}{\gamma}\ln{\left[\frac{\mathrm{cosh}(\gamma Mt)}{\mathrm{cosh}(\gamma Mt_\mathrm{i})}\right]}~~.
 \label{hcr36}
\end{eqnarray}
Here $t_\mathrm{i}$ is the instance of the beginning of inflation when $N = 0$. With $Mt_\mathrm{f} \approx 230$ and in order to have 50 to 65 e-folds of inflation, we find from the above expression that $Mt_\mathrm{i}$ should lie within $Mt_\mathrm{i} = [-10,-26]$. Thus as a whole the inflation starts, i.e when the CMB scale crosses the horizon, at $Mt_\mathrm{i} = [-10,-26]$ and gets an exit at $Mt_\mathrm{f} = 230$, during which, the total e-fold number of the inflationary era becomes $N_\mathrm{f} = [50,65]$.

\begin{figure}[!h]
\begin{center}
 \centering
 \includegraphics[width=3.5in,height=2.5in]{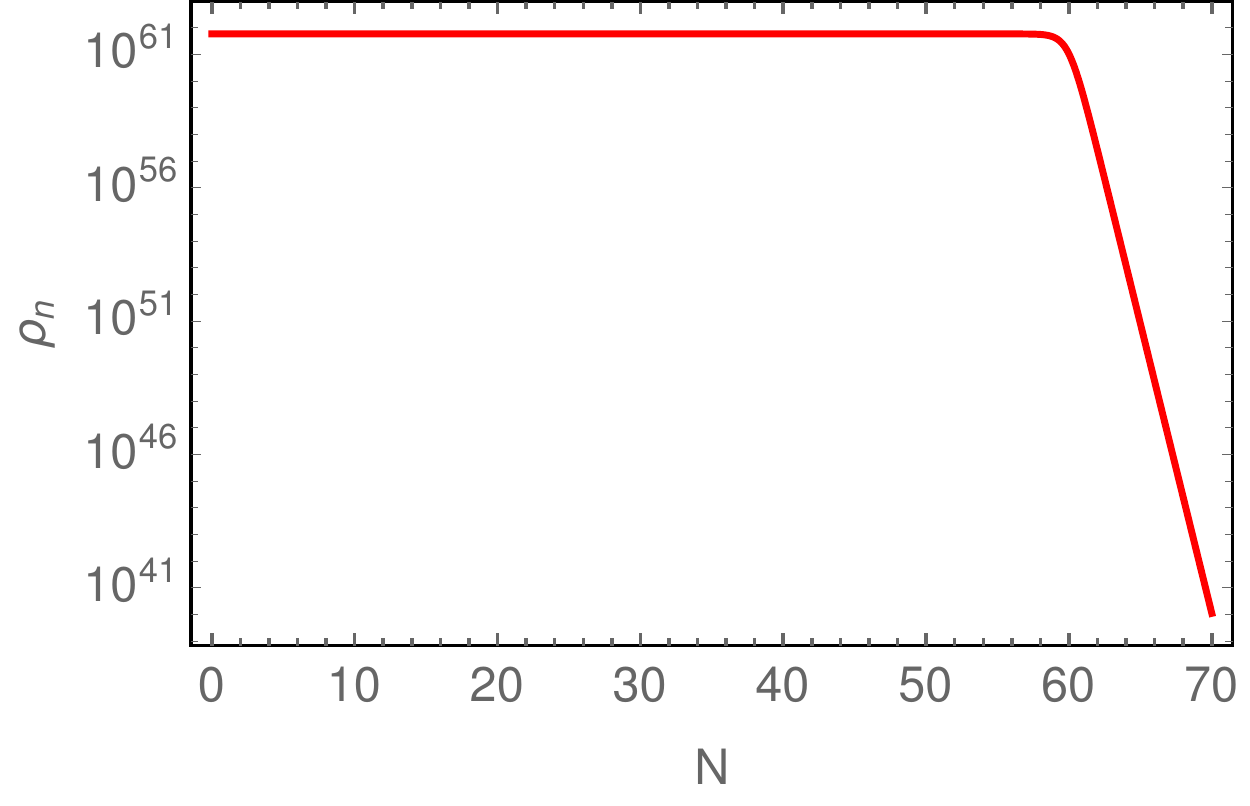}
 \caption{$\rho_\mathrm{n}$ vs. $N$ from Eq.~(\ref{hcr41}) where $\rho_\mathrm{n}$ is in the unit of $\mathrm{GeV}^4$. Here we take $w = \frac{2}{3}$, $N_\mathrm{f} = 60$ and $\rho_\mathrm{n}^{(f)} = 10^{61}\mathrm{GeV}^4$.}
 \label{plot-new2}
\end{center}
\end{figure}

After the end of inflation, the universe needs to enter to a reheating phase or to the standard radiation dominated era in the case of instantaneous reheating. As we have demonstrated that the holographic energy density corresponding to the cut-off $\tilde{L}_\mathrm{IR}$, i.e $\tilde{\rho} = \frac{3}{\kappa^2}\left(\frac{c}{\tilde{L}_\mathrm{IR}}\right)^2$, triggers an inflation during the early universe and a dark energy era during the late time of the universe. Hence in order to have radiation energy in the present scenario, we introduce a new holographic energy density (corresponding to a cut-off $L_\mathrm{n}$), beside the existing $\tilde{\rho}$, and is given by,
\begin{eqnarray}
 \rho_\mathrm{n} = \frac{3}{\kappa^2}\left(\frac{c}{L_\mathrm{n}}\right)^2~~,
 \label{hcr37}
\end{eqnarray}
which is considered to have a coupling with the radiation, i.e around the end of reheating, $\rho_\mathrm{n}$ will eventually decay to the normal radiation. Therefore the total holographic energy density turns out to be,
\begin{eqnarray}
 \rho_\mathrm{T} = \tilde{\rho} + \rho_\mathrm{n} = \frac{3}{\kappa^2}\left(\frac{c}{\tilde{L}_\mathrm{IR}}\right)^2 + \frac{3}{\kappa^2}\left(\frac{c}{L_\mathrm{n}}\right)^2~~,
 \label{hcr38}
\end{eqnarray}
and the effective cut-off ($L_\mathrm{T}$) is given by,
\begin{eqnarray}
 \left(\frac{1}{L_\mathrm{T}}\right)^2 = \left(\frac{1}{\tilde{L}_\mathrm{IR}}\right)^2 + \left(\frac{1}{L_\mathrm{n}}\right)^2~~,
 \label{hcr-rev1}
\end{eqnarray}
where $\tilde{L}_\mathrm{IR}$ is shown in Eq.~(\ref{hcr27}). Similar to $\tilde{\rho}$, the $\rho_\mathrm{n}$ satisfies a conservation equation like
\begin{eqnarray}
 \dot{\rho}_\mathrm{n} + 3H\rho_\mathrm{n}\left(1 + \omega_\mathrm{n}\right) = 0~~,
 \label{hcr39}
\end{eqnarray}
where $\omega_\mathrm{n}$ is the associated EoS parameter of $\rho_\mathrm{n}$ (recall that $\widetilde{\omega}$ is the EoS parameter corresponding to the $\tilde{L}_\mathrm{IR}$, see Eq.~(\ref{hcr33})). Here we would like to mention that the decay rate from $\rho_\mathrm{n}$ to the radiation energy becomes effective only around the end of reheating, and thus $\rho_\mathrm{n}$ safely obeys the above conservation equation. As we will demonstrate below that during the inflation, $\rho_\mathrm{n}$ remains suppressed compared to $\tilde{\rho}$ and thus the inflation is controlled entirely by $\tilde{\rho}$. However after the inflation ends, $\tilde{\rho}$ decreases at a considerably faster rate and lands to the value of the present dark energy density, in particular $\tilde{\rho} \sim 10^{-47}\mathrm{GeV}^4$. As a result, $\rho_\mathrm{n}$ becomes larger over $\tilde{\rho}$ almost one e-fold after the end of inflation and thus dominates the universe's evolution during the reheating stage which ends when the decay rate of $\rho_\mathrm{n}$ becomes comparable to the Hubble rate. As a result, the holographic energy density $\rho_\mathrm{n}$ decays to radiation around the end of reheating, which in turn sets the standard cosmological evolution of the universe. Eventually, the radiation energy density (that falls by $a^{-4}$) becomes less than $\tilde{\rho} \sim 10^{-47}\mathrm{GeV}^{4}$ and then the universe again dominates by $\tilde{\rho}$ which triggers the late time acceleration of the universe. Thus as a whole, the early inflation and the late dark energy era are controlled by $\tilde{\rho}$ with two different energy scales respectively, while the intermediate phase of the universe from the end of inflation to the dark energy era is controlled by $\rho_\mathrm{n}$ and the radiation energy produced from the decay of $\rho_\mathrm{n}$. For the demonstration of such evolution of the universe, we consider a variable EoS parameter corresponding to $\rho_\mathrm{n}$, in particular,
\begin{eqnarray}
 \omega_\mathrm{n}(N) = \left(\frac{1+w}{2}\right)\mathrm{tanh}\left(N - N_\mathrm{f}\right) - \left(\frac{1-w}{2}\right)~~,
 \label{hcr40}
\end{eqnarray}
where $w$ is a constant and recall that $N_\mathrm{f}$ represents the total e-fold number of the inflationary era. According to the above expression, $\omega_\mathrm{n} \approx -1$ for $N < N_\mathrm{f}$ and $\omega_\mathrm{n} \approx w$ during $N > N_\mathrm{f}$ (more-or-less, such behaviour of EoS parameter occurs in alpha attractor scalar-tensor theory with suitable exponent of the scalar potential). Thus the holographic energy density $\rho_\mathrm{n}$ remains almost constant during the inflation, while after the inflation ends, $\rho_\mathrm{n}$ seems to decay by $\propto a^{-3\left(1 + w\right)}$ with the expansion of the universe. In order to get the full evolution of $\rho_\mathrm{n}$, we use the expression of $\omega_\mathrm{n}$ from Eq.~(\ref{hcr40}) to the conservation Eq.~(\ref{hcr39}), and by using $dN = \frac{da}{a}$, we obtain $\rho_\mathrm{n}$ in terms of the e-fold variable as follows:
\begin{eqnarray}
 \rho_\mathrm{n}(N) = \rho_\mathrm{n}^{(\mathrm{f})}\mathrm{exp}\left[-\frac{3}{2}\left(1+w\right)\left\{\left(N - N_\mathrm{f}\right) + \ln{\left[\mathrm{cosh}\left(N - N_\mathrm{f}\right)\right]}\right\}\right]~~,
 \label{hcr41}
\end{eqnarray}
where $\rho_\mathrm{n}^{(\mathrm{f})} = \rho_\mathrm{n}(N_\mathrm{f})$. Therefore the cut-off $L_\mathrm{n}$ corresponding to $\rho_\mathrm{n}$ turns out to be,
\begin{eqnarray}
 \frac{L_\mathrm{n}}{c} = \left(\sqrt{\frac{3}{\kappa^2\rho_\mathrm{n}^{(\mathrm{f})}}}\right)\mathrm{exp}\left[\frac{3}{4}\left(1+w\right)\left\{\left(N - N_\mathrm{f}\right) + \ln{\left[\mathrm{cosh}\left(N - N_\mathrm{f}\right)\right]}\right\}\right]~~.
 \label{hcr42}
\end{eqnarray}
The behaviour of $\rho_\mathrm{n}$ is shown in the Fig.~[\ref{plot-new2}] where we take $w = \frac{2}{3}$ and $N_\mathrm{f} = 60$. The figure clearly demonstrates that $\rho_\mathrm{n}$ remains almost constant during the inflation (i.e during $0 \leq N \leq 60$), while after the inflation ends, $\rho_\mathrm{n}$ seems to decay by $\propto e^{-5N}$ with the e-fold variable. In the Fig.~[\ref{plot-new2}], we also take $\rho_\mathrm{n}^{(\mathrm{f})} = 10^{61}\mathrm{GeV}^4$. The reason for taking such a value of $\rho_\mathrm{n}^{(\mathrm{f})}$ is following: since the inflation is considered to be controlled by $\tilde{\rho}$ which acquires $\tilde{\rho} \sim 10^{66}\mathrm{GeV}^4$ at the beginning of inflation (i.e at $N = 0$) and $\tilde{\rho} \sim 10^{63}\mathrm{GeV}^4$ at the end of inflation (i.e at $N = N_\mathrm{f} = 60$). Thus we take $\rho_\mathrm{n} = 10^{61}\mathrm{GeV}^4$ which remains almost constant during the inflation, so that $\rho_\mathrm{n} \ll \tilde{\rho}$ during $0 \leq N \leq N_\mathrm{f}$ and the inflation gets controlled by $\tilde{\rho}$.

\begin{figure}[!h]
\begin{center}
 \centering
 \includegraphics[width=3.5in,height=2.5in]{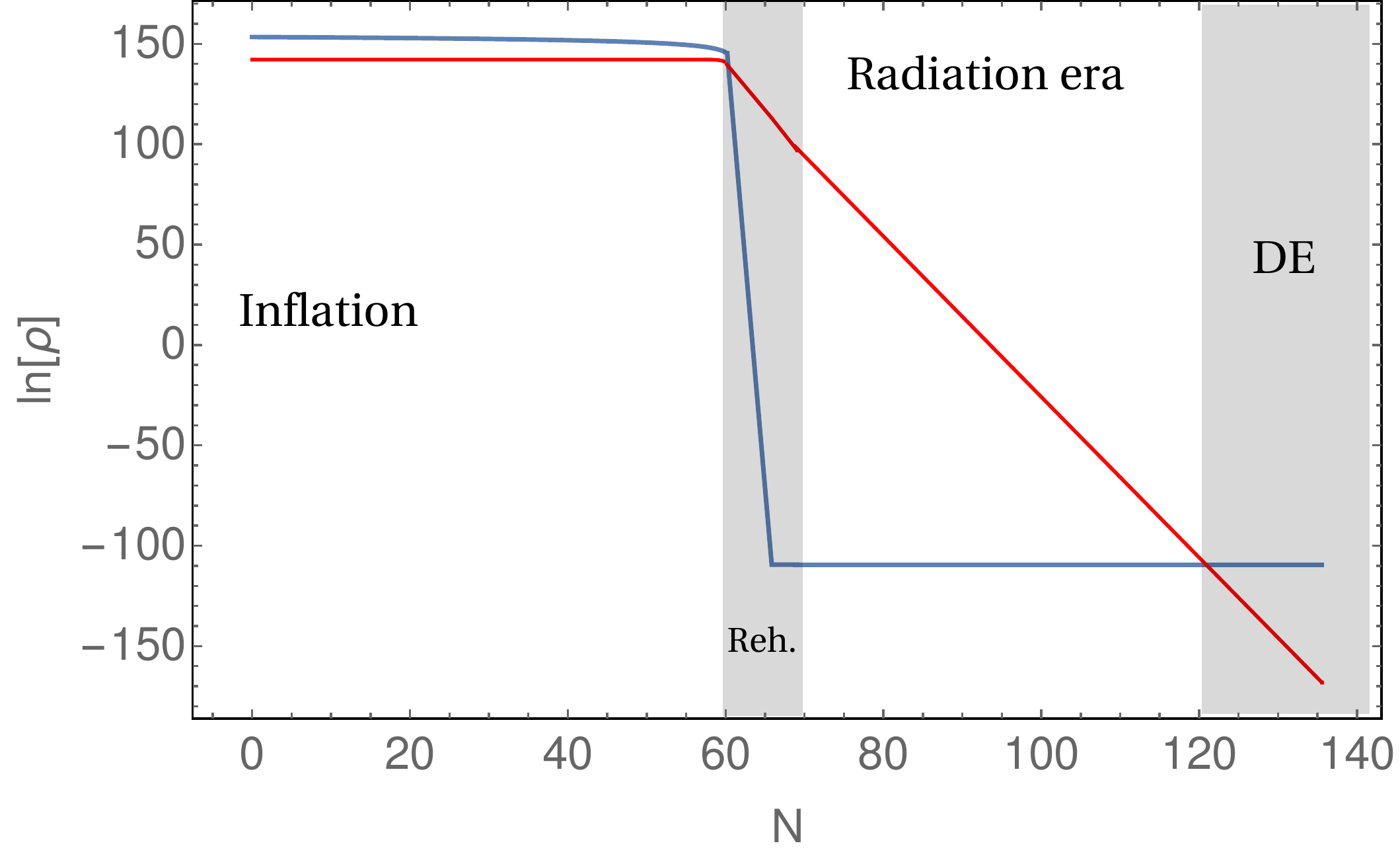}
 \includegraphics[width=3.0in,height=2.5in]{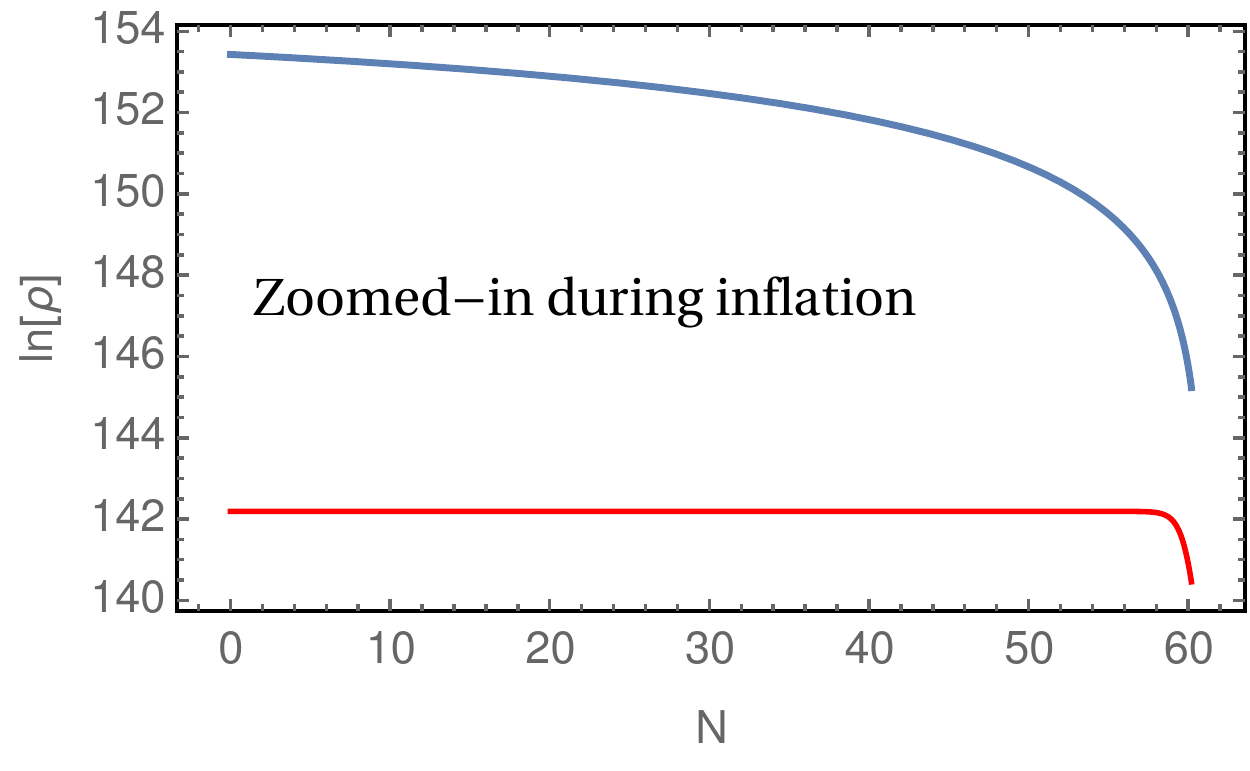}
 \caption{\underline{Left Plot}: $\ln{\tilde{\rho}}$ (the blue curve) and $\ln{\rho_\mathrm{n}}$ (the red curve) with respect to the e-fold variable during the entire cosmological era from the beginning of inflation to the dark energy era. Here the two shaded regions represent the reheating era and the dark energy era respectively, and moreover, the inflation and the radiation era of the universe are represented by the two unshaded regions. The comparisons between $\tilde{\rho}$ and $\rho_\mathrm{n}$ stated in Eq.~(\ref{hcr46}) are clearly evident in this figure. \underline{Right Plot}: The zoomed-in version of the left plot during inflation, i.e for $0 \leq N \leq 60$. We take $w = \frac{2}{3}$, $\rho_\mathrm{n}^{(\mathrm{f})} = 10^{61}\mathrm{GeV}^4$, $\gamma = 0.01$ and $h_0 = 10^{-34}\mathrm{eV}$.}
 \label{plot-new3}
\end{center}
\end{figure}

We now compare the evolutions of $\tilde{\rho}$ and $\rho_\mathrm{n}$ with the expansion of the universe, see Fig.~[\ref{plot-new3}] where the blue and red curves describe the logarithmic scale of $\tilde{\rho}(N)$ and $\rho_\mathrm{n}(N)$ respectively (in particular, the blue curve represents $\ln{\tilde{\rho}}$ and the red curve is for $\ln{\rho_\mathrm{n}}$). The left plot of Fig.[\ref{plot-new3}] shows the two energy densities during the entire cosmological era, i.e from the beginning of inflation to the dark energy era, while the right plot is the zoomed-in version of the left one during the inflation. The figure clearly demonstrates that $\tilde{\rho} \gg \rho_\mathrm{n}$ during the inflation and thus the inflation gets controlled by $\tilde{\rho}$ (corresponding to the cut-off $\tilde{L}_\mathrm{IR}$) which results to a Hubble parameter like Eq.~(\ref{hcr19}). However after the inflation ends, $\tilde{\rho}$ decreases at a faster rate compared to $\rho_\mathrm{n}$, and eventually, $\rho_\mathrm{n}$ dominates over $\tilde{\rho}$ within one e-fold after the end of inflation. Due to the fact that $\rho_\mathrm{n}$ is described by a constant EoS parameter ($w$) during the post inflationary era (see Fig.~[\ref{plot-new2}]), the universe enters to a Kamionkowski like reheating phase during the same.

\begin{figure}[!h]
\begin{center}
 \centering
 \includegraphics[width=3.5in,height=2.5in]{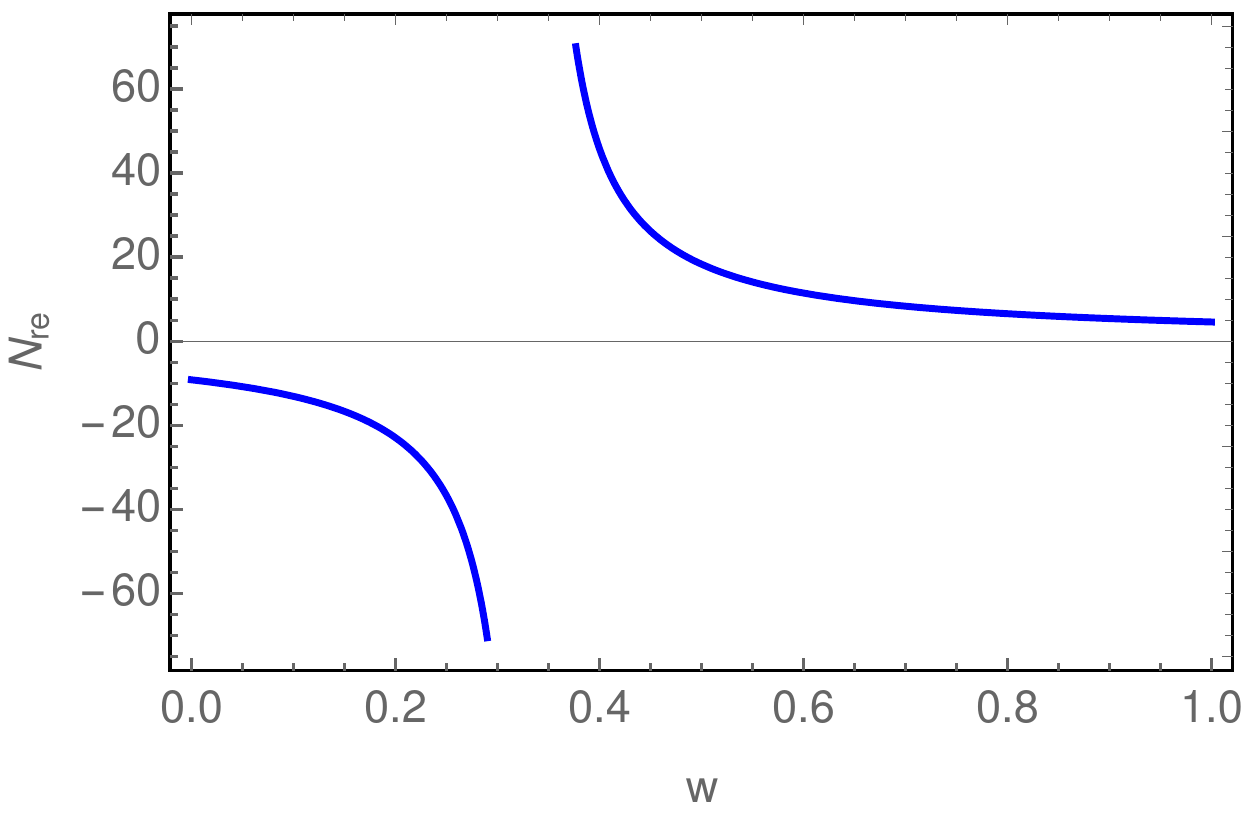}
 \caption{$N_\mathrm{re}$ vs. $w$ from Eq.~(\ref{hcr43}). Here we consider $\gamma = 0.01$ and $h_0 = 10^{-34}\mathrm{ev}$ which are consistent as of Eq.~(\ref{hcr34}).}
 \label{plot-new4}
\end{center}
\end{figure}

In this case, we may write the reheating e-fold number ($N_\mathrm{re}$) and the reheating temperature ($T_\mathrm{re}$) as follows \cite{Cook:2015vqa}:
\begin{eqnarray}
 N_\mathrm{re}&=&\left(\frac{4}{1-3w}\right)\left[61.6 - \ln{\left\{\frac{\left(3H_\mathrm{f}^2M_\mathrm{Pl}^2\right)^{1/4}}{H_\mathrm{i}}\right\}} - N_\mathrm{f}\right]~~,\nonumber\\
 T_\mathrm{re}&=&H_\mathrm{i}\left(\frac{43}{11g_\mathrm{re}}\right)^{1/3}\left(\frac{T_0}{k/a_0}\right)e^{-\left(N_\mathrm{f} + N_\mathrm{re}\right)}~~,
 \label{hcr43}
\end{eqnarray}
where $H_\mathrm{i}$ and $H_\mathrm{f}$ are the Hubble parameter at $N = 0$ and at $N = N_\mathrm{f}$ respectively, $T_0$ is the present temperature of the universe, $k \sim 0.05\mathrm{Mpc}^{-1}$ is the CMB scale and $g_\mathrm{re} \approx 100$ is the relativistic degrees of freedom. From Eq.~(\ref{hcr19}), one can determine $H_\mathrm{i}$ and $H_\mathrm{f}$, and by using these into Eq.~[\ref{hcr43}], we finally obtain the reheating e-fold number in terms of $w$ --- this is shown in Fig.[\ref{plot-new4}]. The figure depicts that in order to be $N_\mathrm{re} > 0$, the EoS parameter of $\rho_\mathrm{n}$ during the reheating stage should be larger than $\frac{1}{3}$, in particular $\frac{1}{3} < w < 1$. For instance, here we consider $w = \frac{2}{3}$ for which the reheating e-fold number comes as $N_\mathrm{re} \approx 10$. Consequently, the reheating temperature becomes $T_\mathrm{re} \approx 1.74\times10^{10}\mathrm{GeV}$ which is indeed safe from the BBN temperature $\sim 10^{-2}\mathrm{GeV}$.

Furthermore the holographic energy density $\rho_\mathrm{n}$ at the end of reheating is given by
\begin{eqnarray}
\rho_\mathrm{n}^{(\mathrm{re})} = \left(\frac{\pi^2g_\mathrm{re}}{30}\right)T_\mathrm{re}^4~~,
\label{hcr44}
\end{eqnarray}
which, due to $T_\mathrm{re} \sim 10^{10}\mathrm{GeV}$, acquires the value as $\rho_\mathrm{n}^{(\mathrm{re})} \sim 10^{40}\mathrm{GeV}^4$ that is also evident in the Fig.[\ref{plot-new3}] (or in the Fig.[\ref{plot-new2}]). At the end of reheating, this amount of $\rho_\mathrm{n}^{(\mathrm{re})}$ decays to radiation which in turn serves the initial energy density in the radiation dominated era. It is clear from Fig.[\ref{plot-new3}] that the radiation coming from $\rho_\mathrm{n}$ still dominates over $\tilde{\rho}$ after the end of reheating and thus sets the radiation dominated era (the standard cosmological evolution) of the universe. We may write the evolution of the radiation energy density (which is the dominating agent during the radiation dominated era) as,
\begin{eqnarray}
 \rho_\mathrm{rad}(N) = \rho_\mathrm{n}^{(\mathrm{re})}~e^{-4\left[N - \left(N_\mathrm{f} + N_\mathrm{re}\right)\right]}~~.
 \label{hcr45}
\end{eqnarray}
Owing to this decaying nature, $\rho_\mathrm{rad}$ eventually becomes comparable to the asymptotic $\tilde{\rho} = 10^{-47}\mathrm{GeV}^4$ which actually sets the present dark energy density of the universe. Eq.~(\ref{hcr45}) immediately indicates that the condition $\rho_\mathrm{rad}(N) = 10^{-47}\mathrm{GeV}^4$ happens nearly at $N = 120$ (i.e almost 50 e-folds after the end of reheating) --- this is consistent with the Fig.[\ref{plot-new3}]. Thus in the present scenario where $w = \frac{2}{3}$ and the other parametric regimes are shown in Eq.~(\ref{hcr34}) --- the inflation occurs during $0 \leq N \lesssim 60$, while the reheating and the radiation dominated era happen during $60 \lesssim N \lesssim 70$ and $70 \lesssim N \lesssim 120$ respectively. Going back to Fig.[\ref{plot-new3}], it is clear that after $N > 120$ (i.e after the radiation dominated era), $\tilde{\rho}$ becomes the dominating agent of the total energy contents. In effect, the universe undergoes through a late time acceleration controlled by a constant energy density $\tilde{\rho} \sim 10^{-47}\mathrm{GeV}^4$. As a whole,
\begin{eqnarray}
 \tilde{\rho}&\gg&\rho_\mathrm{n}~~~~~;~~~~~\mathrm{during~inflation}~~,\nonumber\\
 \rho_\mathrm{n}&\gg&\tilde{\rho}~~~~~~;~~~~~\mathrm{during~reheating}~~,\nonumber\\
 \rho_\mathrm{rad}&\gg&\tilde{\rho}~~~~~~;~~~~~~\mathrm{during~radiation~dominated~era}~~,\nonumber\\
 \tilde{\rho}&\gg&\rho_\mathrm{rad}~~~;~~~~~\mathrm{during~dark~energy~era}~~.
 \label{hcr46}
\end{eqnarray}
As a result, the Hubble parameter during the inflation follows Eq.~(\ref{hcr19}), while during the reheating and during the radiation stages, the Hubble parameter goes as
\begin{eqnarray}
 H(N)&=&H_\mathrm{f}~\mathrm{exp}\left[-\frac{3}{2}(1+w)\left(N - N_\mathrm{f}\right)\right]~~;~~~~~~\mathrm{during~reheating}~,\nonumber\\
 H(N)&=&H_\mathrm{re}~\mathrm{exp}\left[-2\left(N - N_\mathrm{f} - N_\mathrm{re}\right)\right]~~;~~~~~~\mathrm{during~radiation~era}~,
 \label{hcr-rev5}
\end{eqnarray}
respectively. Here $H_\mathrm{re}$ is the Hubble parameter at the end of reheating, and is given by $H_\mathrm{re} = H_\mathrm{f}~\mathrm{e}^{-\frac{3}{2}(1+w)N_\mathrm{re}}$. Finally during the dark energy era, due to the form of $\tilde{L}_\mathrm{IR}$ in Eq.~(\ref{hcr28}), the Hubble parameter becomes a constant $= h_0$. Due to the above expressions of $H(N)$, the dependence of e-fold variable on cosmic time during different epochs are obtained as,
\begin{eqnarray}
 N&=&M\left(t - t_\mathrm{i}\right) + h_0\left(t - t_\mathrm{i}\right) - \frac{1}{\gamma}\ln{\left[\frac{\mathrm{cosh}(\gamma Mt)}{\mathrm{cosh}(\gamma Mt_\mathrm{i})}\right]}~~;~~~~~~\mathrm{during~inflation}~,\nonumber\\
 N&=&N_\mathrm{f} + \frac{2}{3\left(1+w\right)}\ln{\left[\frac{t}{t_\mathrm{f}}\right]}~~;~~~~~~\mathrm{during~reheating}~,\nonumber\\
 N&=&N_\mathrm{f} + N_\mathrm{re} + \frac{1}{2}\ln{\left[\frac{t}{t_\mathrm{re}}\right]}~~;~~~~~~\mathrm{during~radiation~era}~,
 \label{hcr47}
\end{eqnarray}
where $t_\mathrm{i}$, $t_\mathrm{f}$ and $t_\mathrm{re}$ represent the instance of the beginning of inflation, the end of inflation and the end of reheating respectively. Such dependence of $N(t)$ at various cosmological stages have been used in obtaining the Fig.~[\ref{plot-new3}]. As we mentioned earlier that the dark energy starts to dominate the universe nearly at $N \approx 120$, and by using the third expression of Eq.~(\ref{hcr47}), we determine the corresponding cosmic time as $Mt_0 \approx 10^{56}$. With $M = 10^{14}\mathrm{GeV}$ (that sets the inflationary energy scale), we get $t_0 \approx 10^{42}\mathrm{GeV}^{-1} \approx 10\mathrm{By}$ (the conversion $1\mathrm{GeV}^{-1} = 10^{-25}\mathrm{sec}$ may be useful), which suggests that the late acceleration of the universe indeed occurs near the present epoch.

\begin{figure}[!h]
\begin{center}
 \centering
 \includegraphics[width=3.5in,height=2.5in]{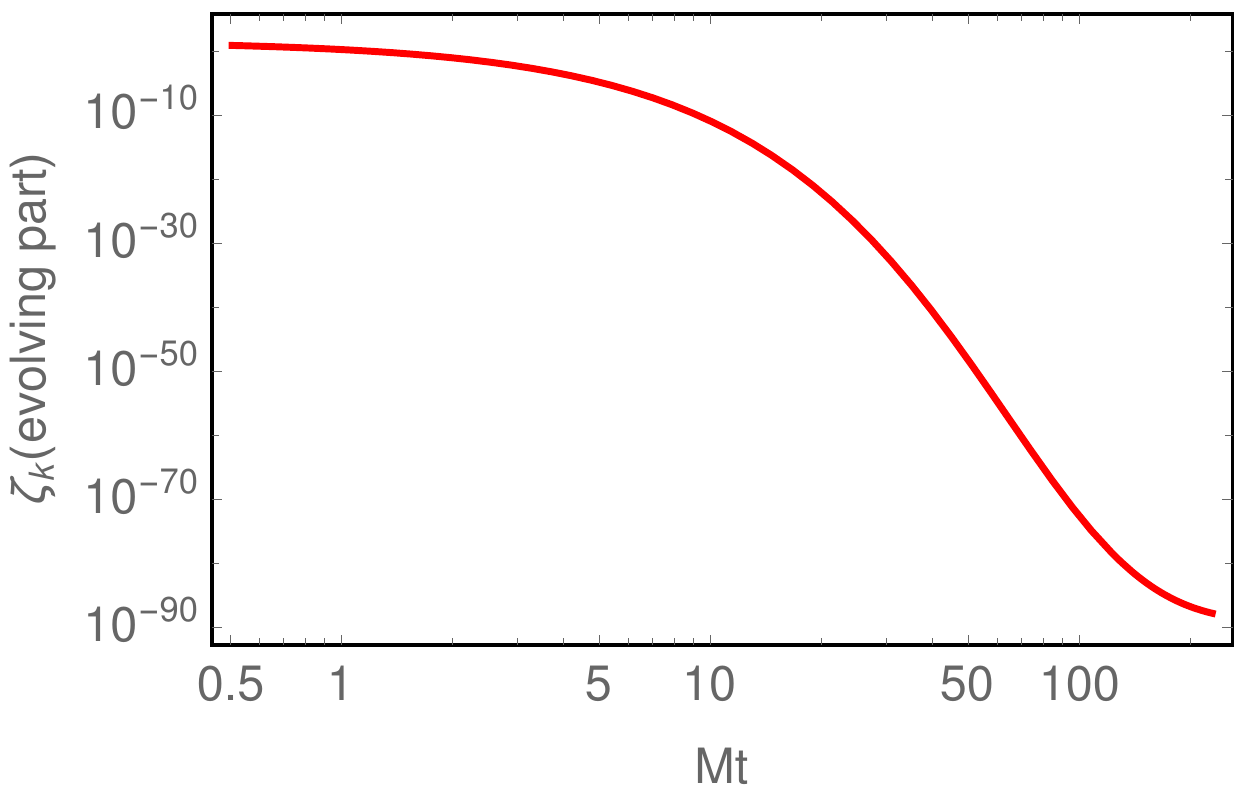}
 \caption{Evolving part of $\zeta_\mathrm{k}(t)$ vs. $t$ during inflation from the first expression of Eq.~(\ref{hcr-rev6}). Here we consider $\gamma = 0.01$ and $M = 10^{14}\mathrm{GeV}$. The figure clearly reveals that the evolving part of $\zeta_\mathrm{k}(t)$ monotonically decays with time and thus the curvature perturbation remains constant in the super-Hubble inflationary regime.}
 \label{plot-new5}
\end{center}
\end{figure}

Besides the background evolutions, the curvature perturbations in the super-Hubble scale are also worthwhile
to address to examine the model's stability. The Fourier mode of primordial curvature perturbation ($\zeta_\mathrm{k}$, with $k$ being the momentum of the Fourier mode) starts from the Bunch-Davies vacuum in the deep sub-Hubble regime (where the perturbation modes of interest lie within the Hubble horizon),
while in the super-Hubble scale, $\zeta_\mathrm{k}(t)$ consists of a constant part and an evolving part as given by,
\begin{align}
\zeta_\mathrm{k}(t) = A_\mathrm{k} + B_\mathrm{k}\int^{t}\frac{dt}{a^3\epsilon_\mathrm{1}}\, ,
\label{per-1}
\end{align}
where $A_\mathrm{k}$, $B_\mathrm{k}$ are constant (with respect to the cosmic time) and $\epsilon_\mathrm{1} = -\dot{H}/H^2$.
The scale factor during inflation behaves as $a(t) \propto \mathrm{e}^{\left(M+h_0\right)t}\mathrm{cosh}^{-\frac{1}{\gamma}}\left(\gamma Mt\right)$, and moreover, $a \propto t^{\frac{2}{3(1+w)}}$ and $a \propto \sqrt{t}$ during the reheating and the radiation era respectively. As a result, the evolving part of $\zeta_\mathrm{k}(t)$ at different cosmological epochs goes as,
\begin{eqnarray}
\int^{t}\frac{dt}{a^3\epsilon_\mathrm{1}}&\sim&\frac{\mathrm{e}^{-3\gamma Mt}\left(1 + \mathrm{e}^{-2\gamma Mt}\right)}{2^{1+\frac{3}{\gamma}}M\gamma\left(3 + \gamma\right)}\mathrm{cosh}^{\frac{3}{\gamma}}\left(\gamma Mt\right)~~;~~~~~~\mathrm{during~inflation}~,\nonumber\\
\int^{t}\frac{dt}{a^3\epsilon_\mathrm{1}}&\sim&t^{-(1-w)/(1+w)}~~;~~~~~~\mathrm{during~reheating}~,\nonumber\\
\int^{t}\frac{dt}{a^3\epsilon_\mathrm{1}}&\sim&t^{-1/2}~~;~~~~~~\mathrm{during~radaiation}~.
\label{hcr-rev6}
\end{eqnarray}
In Fig.~[\ref{plot-new5}], we give the plot of $\int^{t}\frac{dt}{a^3\epsilon_\mathrm{1}}$ during inflation.

Therefore Fig.~[\ref{plot-new5}] along with Eq.~(\ref{hcr-rev6}) indicate that the evolving mode of $\zeta_\mathrm{k}(t)$ monotonically decays with time (as $\gamma > 0$) in the super-Hubble regime from inflation to the radiation dominated era. Hence the curvature perturbation in the present unified scenario becomes constant at super-Hubble scale, i.e.,
\begin{align}
\zeta_\mathrm{k} \approx A_\mathrm{k}\, ,
\label{per-3}
\end{align}
which in turn ensures the stability of the model.

Here we would like to mention that in the present unified scenario, the matter dominated era is absent in-between the radiation and the dark energy epochs. This is because we consider only one decay channel, particularly from $\rho_\mathrm{n}$ to radiation energy, during the reheating stage. In order to introduce the matter dominated stage in the current context, the holographic energy density $\rho_\mathrm{n}$ needs to decay via two channels during the reheating stage: one of these will produce the radiation energy, while the other channel will give the pressureless dust having EoS parameter to be zero. The matter dominated era has its own importance from the fact that the large scale modes re-enter the horizon around the transition of radiation to matter dominated era. Thus unlike to the situation in the radiation era, the large scale perturbation modes remain in the sub-Hubble regime and are not constant with time during the matter dominated epoch. We hope to address these issues (i.e the introduction of matter dominated era in an unified holographic cosmological model and its consequences) at some of our future work.\\

Therefore the holographic model with the cut-off given in Eq.~(\ref{hcr-rev1}) (where $\tilde{L}_\mathrm{IR}$ and $L_\mathrm{n}$ are shown in Eq.~(\ref{hcr28}) and Eq.~(\ref{hcr42}) respectively) proves to unify the universe's evolution from a constant roll inflation to the dark energy era with an intermediate radiation era followed by a Kamionkowski like reheating stage. The inflationary quantities (like the scalar spectral index and tensor-to-scalar ratio) and the present Hubble parameter during the dark energy era lies within the observable regime \cite{Planck:2018jri, Planck:2018vyg}
provided ${\tilde L}_\mathrm{IR}\left(t_\mathrm{h} \right)$, $\gamma$ and $h_0$ satisfy the above-mentioned constraints.

\section{Conclusions}\label{sec-conclusions}

The holographic principle has earned a lot of attention due to its rich cosmological implications for explaining the inflation and the dark energy of the universe.
In the realm of holographic cosmology, the holographic energy density is inversely proportional to the squared infrared cut-off, in particular,
$\rho_\mathrm{hol} \propto \frac{1}{L_\mathrm{IR}^2}$.
However, the fundamental form of the $L_\mathrm{IR}$ is still a debatable topic, and here it deserves mentioning that the most general holographic cut-off
is proposed in \cite{Nojiri:2005pu} where the $L_\mathrm{IR}$ is generalized to depend upon
$L_\mathrm{IR} = L_\mathrm{IR}\!\left(L_\mathrm{p}, \dot L_\mathrm{p}, \ddot L_\mathrm{p}, \cdots, L_\mathrm{f}, \dot L_\mathrm{f},\ddot L_\mathrm{f},
\cdots, a, H, \dot H, \ddot H, \cdots\! \right)$.
Evidently, with such a generalized form of the $L_\mathrm{IR}$, the holographic cosmology becomes phenomenologically richer.

Based on such generalized formalism, we propose a holographic realization of universe's evolution from a constant roll inflation to the dark energy era with an intermediate radiation era followed by a Kamionkowski like reheating stage. The holographic cut-off corresponding to the constant roll inflation depends on the Hubble parameter and its derivatives (up to second order),
and consequently, the cut-off satisfies the equivalent constant roll condition in the holographic scenario.
To examine the viability of the model, we determine the observable quantities like the scalar spectral index ($n_s$) and the tensor-to-scalar ratio ($r$)
in the present holographic inflation, and it turns out that the theoretical expectations of $n_s$ and $r$ become simultaneously compatible
with the Planck data for a suitable value of the model parameter(s).
In this regard, the simultaneous compatibility of $n_s$ and $r$ puts a bound on the holographic cut-off at the instant of horizon crossing.
Such holographic correspondence of constant roll inflation is also extended to the case where the infrared cut-off is corrected by the ultraviolet one which,
during the early universe, may originate from quantum gravity effects.
Due to the appearance of the ultraviolet cut-off, the viable bounds on the infrared cut-off at the instant of horizon crossing modify
compared to the previous case where the ultraviolet correction is absent.
The presence of the ultraviolet correction provides some extra freedom to adjust the model parameters to have a viable
holographic constant roll inflationary scenario. However, these holographic models (without or with ultraviolet correction) are unable to describe the cosmology after the inflation, in particular, the standard cosmological evolution and the dark energy era of the present universe.
In the spirit of this, we propose a modified holographic cut-off which, due to the holographic Friedmann equation,
results in a smoothly unified cosmological scenario from constant roll inflation at an early era to the dark energy era at the late time of the universe. In such unified holographic scenario, the holographic cut-off becomes constant and satisfies the constant roll condition at the early time leading
to successful constant roll inflation, and the cut-off tends to be a different constant at the late time resulting in the dark energy era
of the universe with a lower energy scale compared to that of the inflationary one. The inflation has a graceful exit at a finite time, after which, the universe enters to a Kamionkowski like reheating stage described by a constant EoS parameter of holographic energy density. It turns out that the EoS parameter corresponding to the holographic energy during the reheating stage must be larger than the value $\frac{1}{3}$ in order to have a viable reheating scenario, and moreover, the reheating temperature proves to be safe from the BBN temperature. Around the end of reheating, a portion of the effective holographic energy decays to radiation which in turn sets the standard radiation era of the universe. Regarding the other portion of the effective holographic energy --- it decays at a considerably faster rate after the inflation ends and immediately lands to the present value of the dark energy density almost within 5 e-fold from the end of inflation (see the Fig.[\ref{plot-new3}]), owing to which, it has no role during the radiation dominated era. However due to the fact that the radiation energy density redshifts by $a^{-4}$ and the dark energy density remains almost constant, the universe eventually enters to the dark energy dominated era after a certain time. We have calculated the instance when the radiation energy density becomes comparable to the dark energy density in the present scenario, and it happens at around $t_0 \approx 10\mathrm{By}$ (i.e around the present epoch of the universe). The dark energy EoS parameter tends to the value $= -1$ at a late time, which is however expected because of the constancy
of the late-time Hubble parameter. Consequently, the inflationary quantities (like the scalar spectral index and tensor-to-scalar ratio) and the present Hubble parameter during the dark energy, era prove to be consistent with the observable constraints for suitable ranges of the infrared cut-off
(at the time of horizon crossing during the inflation), the constant roll parameter and the other model parameters. Regarding the evolution of perturbations, it turns out that the curvature perturbation in the present context remains constant (with time) at super-Hubble regime from inflation to the radiation dominated era. This in turn ensures the stability of the unified cosmic model under consideration.

In summary, the generalized holographic formalism proves to be very useful in describing constant roll inflation
during the early universe as well as the unification of constant roll inflation with the late dark energy era of the universe.
However, our understanding of the fundamental cut-off still demands a proper explanation.
We hope that the present work of holographic description of the universe in a unified manner may help in a better understanding of the holographic principle.

\appendix

\section*{{\underline{Appendix}}: Holographic cut-offs in terms of either particle horizon or future horizon}\label{appendix}

Using Eq.~(\ref{HLL}) into Eq.~(\ref{hcr1}), one may obtain the $L^{(1)}_\mathrm{IR}$ either in terms of particle horizon ($L_\mathrm{p}$)
and its derivatives or in terms of the future horizon ($L_\mathrm{f}$) and its derivatives. They are given by:
\begin{align}
L^{(1)}_\mathrm{IR} = 2c\beta\left(\frac{\ddot{L}_\mathrm{p}}{L_\mathrm{p}} - \frac{\left(\dot{L}_\mathrm{p}\right)^2}{{L_\mathrm{p}}^2}
+ \frac{\dot{L}_\mathrm{p}}{{L_\mathrm{p}}^2}\right)\left(\frac{\dddot{L}_\mathrm{p}}{L_\mathrm{p}} - \frac{3\dot{L}_\mathrm{p}\ddot{L}_\mathrm{p}}{{L_\mathrm{p}}^2}
+ \frac{2\left(\dot{L}_\mathrm{p}\right)^3}{{L_\mathrm{p}}^3} + \frac{\ddot{L}_\mathrm{p}}{{L_\mathrm{p}}^2} - \frac{2\left(\dot{L}_\mathrm{p}\right)^2}{{L_\mathrm{p}}^3}\right)^{-1}\, ,
\end{align}
in terms of $L_\mathrm{p}$ and its derivatives, or similarly,
\begin{align}
L^{(1)}_\mathrm{IR} = 2c\beta\left(\frac{\ddot{L}_\mathrm{f}}{L_\mathrm{f}} - \frac{\left(\dot{L}_\mathrm{f}\right)^2}{{L_\mathrm{f}}^2}
 - \frac{\dot{L}_\mathrm{f}}{{L_\mathrm{f}}^2}\right)\left(\frac{\dddot{L}_\mathrm{f}}{L_\mathrm{f}} - \frac{3\dot{L}_\mathrm{f}\ddot{L}_\mathrm{f}}{{L_\mathrm{f}}^2}
+ \frac{2\left(\dot{L}_\mathrm{f}\right)^3}{{L_\mathrm{f}}^3} - \frac{\ddot{L}_\mathrm{f}}{{L_\mathrm{f}}^2} + \frac{2\left(\dot{L}_\mathrm{f}\right)^2}{{L_\mathrm{f}}^3}\right)^{-1}\, ,
\end{align}
in terms of $L_\mathrm{f}$ and its derivatives.
Furthermore from Eq.~(\ref{HLL}) and Eq.~(\ref{hcr2}), we obtain $L^{(2)}_\mathrm{IR}$ in terms of particle horizon and its derivatives as follows:
\begin{align}
L^{(2)}_\mathrm{IR} = c\beta\left(\frac{\dot{L}_\mathrm{p}}{L_\mathrm{p}} - \frac{1}{L_\mathrm{p}}\right)\left(\frac{\ddot{L}_\mathrm{p}}{L_\mathrm{p}}
 - \frac{\left(\dot{L}_\mathrm{p}\right)^2}{{L_\mathrm{p}}^2} + \frac{\dot{L}_\mathrm{p}}{{L_\mathrm{p}}^2} + \beta M^2\right)^{-1}\, ,
\end{align}
and moreover,
\begin{align}
L^{(2)}_\mathrm{IR} = c\beta\left(\frac{\dot{L}_\mathrm{f}}{L_\mathrm{f}} + \frac{1}{L_\mathrm{f}}\right)\left(\frac{\ddot{L}_\mathrm{f}}{L_\mathrm{f}}
 - \frac{\left(\dot{L}_\mathrm{f}\right)^2}{{L_\mathrm{f}}^2} - \frac{\dot{L}_\mathrm{f}}{{L_\mathrm{f}}^2} + \beta M^2\right)^{-1}\, ,
\end{align}
in terms of $L_\mathrm{f}$ and its derivatives.
Thus the above four equations provide our desired results for this section.

\section*{Acknowledgments}

This work is supported in part by MICINN (Spain), project PID2019-104397GB-I00 and JSPS fellowship S23013 (SDO).

\end{document}